# Intelligent Computing Social Modeling and Methodological Innovations in Political Science in the Era of Large Language Models

**Zhenyu Wang[1] Dequan Wang[3] Yi Xu[2*] Lingfeng Zhou[3*] Yiqi Zhou[4]**

[1]Shanghai Academy of Social Sciences [2]Nanjing University [3]Shanghai Jiao Tong University [4]Shanghai Institutes for International Studies

lucienw@sass.org.cn xuyi@smail.nju.edu.cn zhoulingfeng@sjtu.edu.cn

**Abstract:** The recent surge in artificial intelligence, exemplified by large language models (LLMs), has created both opportunities and challenges for methodological innovation in political science. This has sparked discussions about a potential paradigm shift in the social sciences. Nevertheless, how can we comprehensively understand the influence of LLMs on the production and transformation of knowledge in the social sciences from an integrated perspective that encompasses both technology and methodology? What are the specific applications of LLMs and the representative innovative methods in political science research? This paper proposes the Intelligent Computing Social Modeling (ICSM) framework as a systematic approach to addressing these issues through a comprehensive analysis of LLM mechanisms. ICSM capitalizes on the capabilities of LLMs in idea synthesis and action simulation, thereby facilitating intellectual exploration in political science through "simulated society construction" and "simulation validation." We empirically demonstrate the operational pathways and methodological advantages of ICSM through a simulation of the U.S. presidential election. By integrating traditional social science paradigms, ICSM enhances the quantitative paradigm's ability to leverage big data for assessing variable impacts while also providing the qualitative paradigm with empirical evidence for mechanism discovery at the individual level. This methodology achieves a balance between interpretability and predictive accuracy in social science research. Our framework offers practical tools for analyzing complex social phenomena using large language models.

## 1 Introduction

Artificial intelligence (AI) is becoming increasingly influential across various domains of human society, profoundly shaping research methods and paradigms. The rapid advancement of generative AI, exemplified by large language models (LLMs) such as OpenAI's ChatGPT, has has significantly enhanced AI's ability to model and simulate the real world. This progress heralds the imminent advent of strong AI, bringing unprecedented transformations and opportunities to scientific research. Some scholars have emphasized the role of LLMs in natural science research and discovery, particularly in applications such as protein folding [74], while others have highlighted their expanding potential in the social sciences. In particular, Transformer-based models trained on large textual datasets are increasingly capable of simulating human-like

* Corresponding author. Address of correspondence:School of Government, Nanjing University, Shengda Building, 163 Xianlin Avenue, Nanjing 210023, China.

responses and behaviors [29].

The integration of AI and social sciences can be categorized into two main areas. The first involves leveraging AI to enhance existing paradigms in social science research, while the second examines AI as a central subject of social science inquiry [81][①]. On the one hand, AI serves as a crucial tool that enables researchers to efficiently conduct tasks such as literature reviews, hypothesis formulation, data collection, and social surveys. On the other hand, AI is increasingly employed in social science research to simulate social actors, analyze behavioral patterns, and address complex issues by leveraging its cognitive, reasoning, and linguistic capabilities. While there is a consensus that AI is a valuable tool for advancing social science research, whether AI should be regarded as a core research subject remains a topic of debate. This discussion primarily revolves around the following key points.

First is the relationship between LLMs and existing big data research. Some discussions frame LLMs as an extension of big data research methodologies. For instance, some scholars characterize social science research in the AI era as inherently data-intensive, emphasizing LLMs' ability to expand the scope of data from structured forms to multimodal, high-density datasets [21, 53]. While these perspectives accurately capture the data-driven foundation of LLMs, they often overlook the models' logical reasoning and content generation capabilities. The big data approach summarizes existing data and identifies patterns, whereas LLMs go beyond this by generating coherent natural language content. The differences in training costs, complexity, and encapsulation are considerable: while the big data approach focuses on specific machine learning tasks, LLMs are designed to preserve human-like logic and language expression in complex scenarios.

The second issue concerns the potential impact of generative AI on theoretical inquiry in the social sciences. Some scholars argue that data generated by LLMs may be unreliable and should not be used for future social science analysis [47]. The generation of data by LLMs presents several challenges, including the diversity of the data produced, the potential for bias [85], and the occurrence of model hallucination [86]. Hu Anning notes that since generative AI heavily relies on data-related instructions, systematic biases in the data will inevitably lead to biased outputs [36]. Without precise descriptions, AI struggles to generate an accurate representation of the subject [7]. Conversely, an alternative perspective suggests that AI and LLM-generated data can serve as an analytical foundation, potentially driving paradigm shifts in social science research. For example, some research indicates that with the capacity for automated analysis, LLMs have the potential to transcend traditional paradigms, facilitating interdisciplinary integration and knowledge expansion in social sciences [88].

The existing literature underscores the pivotal role of LLMs in advancing research efforts. However, their potential contributions go far beyond this domain. By harnessing the growing power of pretrained models, LLMs are poised to become integral to the core processes of

① Some scholars summarized these distinctions as "AI for Social Science" and "Social Science of AI." (Xu et al. 2024)

knowledge exploration in the social sciences. For this integration to occur, a systematic epistemological and methodological framework is required to deeply embed LLMs into social science inquiry.

This paper will be structured into three main sections to achieve this explorations. The second section examines the interrelationship between the technical characteristics of LLMs and the evolution of social science research methodologies, elucidating the mechanisms through which LLMs facilitate social science research. The third section provides a detailed elaboration of the Intelligent Computing Social Modeling (ICSM) method, outlining the standardized operational procedures that underpin it. The fourth section will conclude with a discussion of the practical applications of the method. It highlights LLMs' contributions and extensive applications in theoretical exploration, focusing on classic theoretical scenarios and specific research questions.

## 2 Key Mechanisms of LLMs Empowering Political Science Research Methods

The effectiveness of research methods depends on their ability to accurately and consistently capture social processes. Social science research primarily focuses on ideas, actions, and outcomes [71]. A comprehensive theoretical narrative requires the explicit or implicit delineation of these three elements, as their interactions inform social processes and give rise to many social phenomena. Accordingly, different research methods can be regarded as distinct approaches to representing social processes within the "theoretical space." The objective is to ensure that theoretical frameworks accurately capture real-world social processes, whether using "counterfactuals" in causal inference, "pattern recognition" in predictive studies, or "rule-based" models in simulation research. Precise observation and depiction of ideas and actions are essential for effective representation in social science. Without such precision, no social science model can provide valid evidence. The ICSM method incorporates LLMs into social science research in a manner that is closely aligned with these considerations, thus demonstrating a practical approach to integrating these novel tools into established research practices. We emphasize that LLMs' capacity for idea synthesis and action simulation underscores their significant potential to advance social science research.

### 2.1 Idea synthesis

LLMs, trained on extensive datasets, serve as an unparalleled repository of human ideas and perceptions at the most granular level. They represent a groundbreaking technological advancement that enables direct analysis and computation of non-numerical data, including language and images. Consequently, the models encode detailed identity-related information about real world actors. Using computational simulation experiments[①] based on this data facilitates a more refined characterization of specific actors than is possible with traditional research methods.

---

[①] In the context of ICSM, experiment refers to a structured research process that uses computational simulations to explore, verify, or compare outcomes under specific settings.The experiment is designed around a specific simulation task, with researchers constructing the simulation process by defining the agents' identities and actions. The effectiveness of different experimental settings is assessed by comparing them to benchmarks. In this paper, ICSM experiments are presented in two forms: one involves experiments under different settings, and the other involves repeated experiments under the same setting. They serve different methodological purposes.

In conventional inductive methodologies, social actors are characterized by aggregating data from interviews and surveys. Researchers employ various methods to extract insights from empirical materials. Consequently, the quantity and utilization of information determine the method's efficacy in portraying the actors. LLMs exhibit notable advantages in both areas. To illustrate, Google's publicly-announced model, Palm2, was trained on 3.6 trillion tokens.[①] NVIDIA's Nemotron-4 employed 9 trillion tokens. These language models are trained on large-scale textual datasets from academic websites, online encyclopedias, social media platforms, and forums. These datasets encompass a vast amount of textual data. The substantial data about human language and society embedded within these texts serves as the foundational material for machine learning.

Furthermore, the Scaling Law observed in LLM training ensures effective information transmission. It describes the power-law relationship between model performance and factors such as model size, dataset size, and the amount of computing used during training [41]. Expanding both models and datasets results in enhanced model performance. The technical architecture of LLM systems enables them to "decode" embedded knowledge, and their capacity to learn intricate details within data increases in proportion to the size of the model.

Expanding from purely language-based models to multimodal models, artificial intelligence will gain a deeper and more nuanced understanding of real-world contexts in the future. The integration of multimodal data ensures the preservation of the content of the information, regardless of its form. The data and parameter scale that underpin LLM models endow them with exceptional capabilities in natural language processing. Moreover, the Transformer architecture has also exhibited excellent performance in processing multimodal information. This engineering advancement is reflected in academic research and industry applications, as evidenced by the development of multimodal large models such as Sora. In the context of traditional inductive approaches, it is evident that no single method can effectively handle multimodal information. Data modeling frequently results in the loss or compression of textual and image data. Case analysis employs selective information, reflecting the subjective viewpoints inherent to this approach. In light of the considerations above, it can be concluded that “computable” multimodal information signifies that LLMs can transform real-world information to the analytical stage in a more holistic manner.

Considering these technical capabilities, LLMs present a novel opportunity for a more thorough and nuanced characterization of social actors. LLMs can generate “silicon samples” that exhibit analogous political characteristics to those observed in real-world actors. These samples are derived from LLMs' vast training data and world knowledge, they can understand and respond to complex political ideas. This enables the effective simulation of political behaviors. LLMs

---

① Considering the conversion between tokens and English words implies that the model was trained on 2.77 trillion English words.

contain information that far exceeds that of any conventional database.① These LLMs could complement traditional case analysis by processing multimodal information. By employing appropriate "prompt engineering," researchers could construct representative samples and utilize them as survey respondents.② Furthermore, the creation of intelligent agents allows for synthesizing demographic data, socio-economic data, and political archives, generating actors at varying scales (e.g., social organizations, political parties, and nation-states). This enables the analysis of specific interactions and decision-making processes.③

**2.2 Action Simulation**

One of the primary tasks within the field of political science, and more broadly within the social sciences, is to explain the rationale behind a given action. From a theoretical standpoint, an understanding of the interplay between human cognition and the external environment can be achieved by examining human behavior. Furthermore, the interpretation of political phenomena at various levels and scales inevitably necessitates the integration of these phenomena through actions [69]. LLMs, functioning as implicit computational models of human action and as social context framers capable of constructing complex external environments, offer a novel approach to overcoming the "simplification dilemma" in action simulation.④

Methodologically, behavior modeling is constrained by significant limitations. The objective of studies employing this method is to undertake empirical and mathematical observation, description, and analysis of political behavior to identify verifiable laws of human behavior. This approach is designed to enhance the reliability and objectivity of political science [22]. To uncover the motives behind human actions, political science frequently deploys formal modeling and experimental techniques that closely mirror the methodologies employed in natural science research. Formal modeling provides mathematically rigorous tools for explaining actions, while experiments offer empirically robust evidence with high internal validity and causal reliability [52]. Together, these methods provide a scientific framework for the precise formulation of theories and a rigorous examination of hypotheses. However, both methods have significant limitations in requiring highly simplified action patterns and external environments, which significantly restrict the external validity of research findings.

Following extensive discussions on AI reasoning and cognitive abilities, researchers have

① The GDelt database, widely used in large-scale data research, only contains one ten-millionth of GPT-4's training data volume.

② The prompt engineering here focuses on assigning attributes such as ethnicity, socioeconomic status, and political preferences to the LLM through prompts. Then researchers engage in interactive dialogue to query LLM-based agents' political views, which will ultimately be randomized and compared with real-world results according to the proportion of demographic data (Sun et al. 2024).

③ The highest-level scenario involves nation-states. The logic here is to input decision-making variables such as population, resources, and strategic objectives into the LLM and then ask it to make its final choices among decision options through interactive dialogue (Hua et al. 2023).

④ The argument regarding LLM as implicit computation models of human action can be found in (Horton 2023). This paper emphasizes that the text on which LLMs are trained reflects human actions' thinking and reasoning processes. It also highlights that LLMs can acquire attributes, information, and preference data to make human-like decisions in simulated scenarios. The discussion on LLM as social context framers can be referenced in (Lorè and Heydari 2023). This paper simulated five multi-level game environments (from diplomatic relations to casual friendships) to demonstrate that contextual prompts can flexibly set social scenarios.

demonstrated that LLMs can perform "reasoning-based actions" when prompted in a specific manner. The essential techniques of prompt engineering include In-Context Learning (ICL) and Chain of Thought (CoT) prompting, and their significance in simulation studies has been well-proven.

CoT and ICL are integral to prompt engineering, which optimizes responses. Prompt engineering encompasses designing and refining prompt input to LLMs to guide them in generating more accurate and valuable responses [77]. CoT facilitates the model's acquisition of a more profound comprehension of the problem, thereby enhancing the precision of the output. Meanwhile, ICL entails incorporating sufficient background information within the prompts, frequently ascribing particular roles or contexts to the model. This allows the model to accurately comprehend the contexts or conditions in which the problem is situated [79]. ICL may include relevant definitions, previous information, or other details that facilitate the model's comprehension of the task more accurately. The incorporation of well-designed prompts has been demonstrated to enhance the model's understanding of the issues at hand and the relevance of its outputs. Optimization can be said to occur in two distinct directions. The ICL approach contextualizes the model's responses based on the roles and contexts to which they are assigned. At the same time, the CoT reasoning process enhances the model's ability to understand and handle complex tasks [75]. These technologies have made LLMs an increasingly preferred tool for natural language interaction and a highly effective tool for social simulation [59]. For example, Argyle et al. employed prompt engineering to construct LLM-based agents directly and simulated the outcomes of U.S. presidential elections based on their responses [4]. The combination of CoT and ICL allows for effective simulation of interactions between actors [82]. Such intelligent agent simulations are not confined to virtual scenarios [37]. The simulation approach is not limited to simple tasks: researchers have explored how external models of world knowledge can enhance the optimization effects of CoT reasoning [31]. In conclusion, constructing intelligent agents based on prompt engineering represents a significant avenue for applying LLMs in social science research. By employing suitable prompts, researchers can generate "human-like behavior" in simulated samples within simulation experiments, thereby enabling the tracing of "human-like reasoning" embedded in these behaviors.

Firstly, LLMs can effectively simulate at least eight types of behavior from a technical perspective, given their high consistency with human behavior patterns. These eight types of behavior are as follows: natural language-based interaction, knowledge acquisition and memory, reasoning and planning, learning and adaptation, social interaction, personalization and emotional expression, self-improvement and self-adjustment, and tool use and innovation [80]. The combination of these eight types of behavior enables the simulation of LLMs to encompass the behavioral modes observed in mainstream formal models, including those about rational choice, bounded rationality, game theory, and psychological processes [57, 66, 8, 6]. Furthermore, the advent of LLMs has brought about a paradigm shift in the simulation of social scenarios in action research. This is due to three significant factors: the diversification of operational approaches, the enhancement of granular detail, and the broadening of scenario scope. Ideally, LLMs can textualize intricate social scenarios and game rules and draw upon authentic data to inform the simulation of social context. This provides the capability to conduct experimental simulations without constraints on duration, rounds, number of participants, or modes of participation, thus enabling researchers to overcome resource limitations in action research readily.① The high flexibility of this approach allows researchers to replicate critical scenarios in action research, including those involving cooperation and betrayal,② economic behavior [64], collective decision-making [40], moral judgment, and cognitive psychology scenarios.③ Multi-round games and multi-agent negotiations are also noteworthy topics [24, 45, 1]. Notably, research on strategic behavior indicates that LLMs can be utilized to simulate actions across various levels, including states, social organizations, and individuals [50].

## 3.Intelligent Computing Social Modeling: A New Approach in Political Science Research Driven by Generative AI

### 3.1 Method Definition and Philosophical Foundation

The rapid evolution of generative AI enables the integration of this technology into computational simulations, thereby facilitating the analysis and forecasting of dynamic social outcomes in political science. This has resulted in a novel approach to political science research. ICSM is a modeling methodology that employs LLMs to construct actors, social contexts, and action patterns. It synthesizes ideas, actions, and outcomes by leveraging data produced by generative agents for an integrated computational simulation. At the level of the actor, the ICSM method generates a cohort of agents that mirror real-world demographic profiles. Each agent acquires political perceptions from LLMs, aligning with the actual distribution within a defined temporal and spatial context. At the societal level, ICSM furnishes agents with environmental data via LLMs, utilizing manually processed natural language inputs or multimodal data obtained through In-Context Learning (ICL). Regarding action patterns, ICSM guides simulated agents in undertaking particular actions by providing information to LLMs.④ The behaviors exhibited by

① In existing research, multiple rounds of prompts are often used to assign game rules or background information to the LLM in text format, along with the agents' identity attributes and behavioral preferences. This is followed by multi-round question-answering experiments. Researchers can also integrate frameworks such as AutoGen or CAMEL, which directly generate different types of agents using the LLM and instruct them to conduct conversations and interactions under specific rules. Researchers can also use APIs to input real-world data for the agents and add intervention strategies, such as an information dissemination model that affects information distribution to simulate how humans process external environments more realistically. For example, Zhao et al. used restaurant reviews from the Bookings as a social environment, leveraging LLMs to simulate interactions between restaurant competitors and customers (2023). The study employed multi-round interaction and mutual imitation to explore business competition interactions, yielding results aligned with economic theory.

② Examples of cooperation games include "Guess 2/3 of the Average," "El Farol Bar problem," and "Divide the Dollar." Betrayal games include the "Public Goods Game," "Diner's Dilemma," and "Sealed-Bid Auction." (Huang et al. 2023)

③ Examples include the Dictator game and the Ultimatum game (Dillion et al. 2023).

④ The input here can be based on natural language prompts or execution of R or Python code. The instructions given could be a specific action (e.g., requesting the agent to provide voting choices), a set of action rules (e.g., asking the agent to spread political ideas according to the given proportion), or an action goal (e.g., asking the agent to gain more support through interaction).

these agents result from cognitive perception and decision-making processes undertaken by the language models.[①] ICSM employs computational data from generative AI to examine meso and macro-level political science issues. It develops an LLM-based intelligent simulated society closely resembling real-world demographics, social contexts, and behavioral patterns.

The ontological basis of ICSM is grounded in Karl Popper's "Three Worlds" theory [61], which posits a tripartite and open-causal ontology consisting of three coherent worlds: the physical world (World 1), the mental world (World 2), and the knowledge world (World 3). World 1 is constituted by all physical bodies, states, and processes. World 2 comprises mental states or psychological processes, representing the world of conscious experience. World 3 is the world of the products of the human mind, including all knowledge, scientific theories, and cultures created by humans. In Popper's framework, the physical world serves as the foundation for the mental and knowledge worlds, while the mental world acts as an intermediary between the physical and knowledge worlds. The interaction between the physical and mental worlds gives rise to the formation of the knowledge world, which, in turn, exerts influence on the physical world via the mental world. All three worlds are, in fact, real and interact dynamically [61][②].

The advent of ICSM can be attributed to the fact that LLMs furnish researchers with the technical apparatus to establish a coherent nexus between the three worlds. From the perspective of the evolution of artificial intelligence technology, LLMs represent a shift in machine learning from pattern recognition in specific datasets to probabilistic generation based on ultra-large-scale datasets.[③] In this sense, the development goal of LLMs is to create a "simulator" that synchronizes the evolution of the three worlds. Firstly, the evolution of LLMs fundamentally depends on the expansion and updating of parameter scales and training datasets. This enables LLMs to reflect the changing dynamics in the physical world [78]. Secondly, the performance of LLMs is gauged by the extent to which they mirror human cognitive processes. This entails assessing the precision with which their generative capabilities emulate the operations of the mental realm [15] . In conclusion, the content generated by LLMs contributes to the knowledge world, which can be applied to the physical world through interaction with the mental world.[④] By precisely simulating the interactions among the three worlds, LLMs enhance the value and advantages of the ICSM method, thereby driving innovative developments in political science

---

① Agent-based modeling using LLMs can be referenced in the "Generative Agent-Based Modeling" (GABM) method (Ghaffarzadegan et al. 2023).

② The concept of the "three worlds" is widely used as an ontological or philosophical basis in research related to computational simulation and social simulation (Yang et al. 2023; Wang et al. 2016).

③ Classic machine learning methods for pattern recognition include logistic regression, support vector machines (SVM), and feedforward neural networks. These methods primarily focus on classification tasks, identifying latent patterns in unstructured data such as text or images by creating algorithmic models. Generative models, such as LLMs, operate differently. These models assume that data exists based on a probabilistic distribution. The approach's core involves training models on a massive dataset to generate a distribution that closely resembles the original data. Based on this generated distribution, the model calculates the probability of a target variable under given input conditions, ultimately producing free-form outputs such as text, images, and videos (Harshvardhan et al. 2020).

④ The widespread applications of LLMs in fields like engineering, medicine, education, and finance can be referenced in (Hadi et al. 2023).

research methodologies.

**3.2 Operation Workflow：Simulated Society Construction & Simulation Validation**

To fully leverage the technological advantages of LLMs to empower political science research, the operational workflow of ICSM have been designed around two cycles corresponding to the "three worlds." The first is the *Simulated Society Construction* pathway, which corresponds to the cycle of "physical world - mental world - knowledge world - physical world." This pathway consists of four steps: *Construction of AI Agent, Action Simulation, Intelligent Computing Simulation, and Generated Data Analysis*, aiming to construct a simulated society that can generate data for research. The second is the simulation validation pathway, which corresponds to the cycle of "knowledge world - mental world - physical world - knowledge world." This path also consists of four steps: *Validity Benchmark Setting, Simulated Reliability Test, Simulated Validity Assessment, and Knowledge Discovery Validation*, which correspond to the *Simulated Society Construction*, aiming to ensure the reliability of generated data in knowledge discovery tasks. The operational workflow and application logic of the two principal pathways are shown in the Figure.1, and each step of the two paths will be introduced separately as follows.

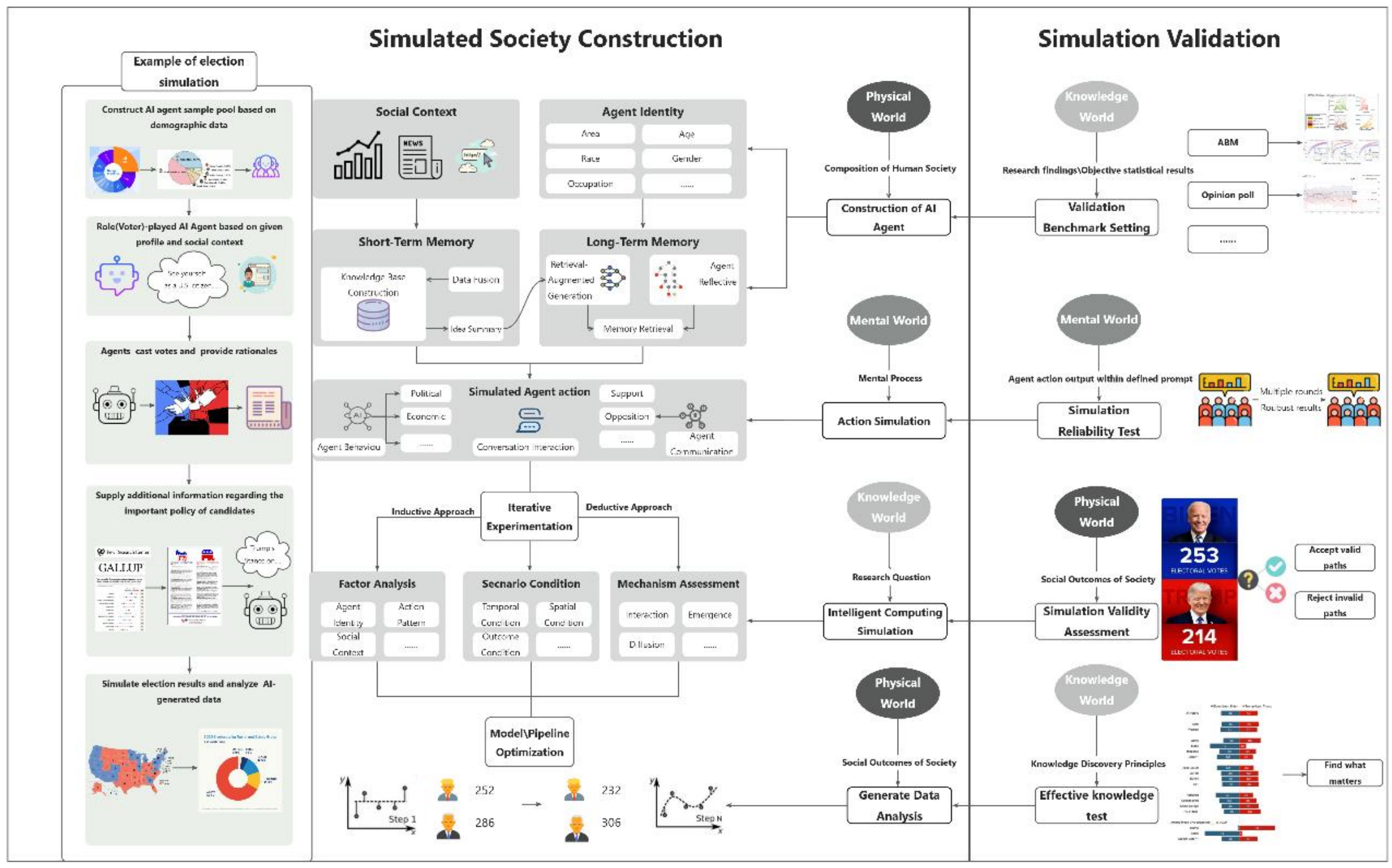

**Fig. 1** Operational workflow and application logic of the ICSM method

**3.3.1 Simulated Society Construction**

To study any social outcome, ICSM must complete the following four steps along the simulated society construction path：

Step.1 Construction of AI Agent

The initial step of the ICSM entails the Construction of AI agent samples. This step is designed to create AI agent samples that accurately reflects the actual demographic composition of the physical world. We use the term sample to denote our intention to construct a subset of the social population that replicates the real-world distribution of key social characteristics, ensuring its representativeness of the broader population. The resulting sample will serve a purpose similar to that of survey methods. In particular, this phase is concerned with the construction of a cohort of agents that reflect the demographic structure of reality based on census data. Each agent is assigned an identity corresponding to the demographic characteristics observed in reality. This identity is then populated with perceptions drawn from the LLM's extensive data repository on the physical world.

Step.2 Action Simulation

Action Simulation is the second step in which actions for agents that correspond to cognition in the mental world are established. This aligns with action-focused formal models. This step entails prompting the agents to perform actions that can directly influence social outcomes, such as voting. Prompt engineering techniques that are currently in use in mainstream research include zero-shot and few-shot prompting, CoT reasoning, retrieval-augmented generation (RAG), and Automatic Reasoning and Tool Use(ART). Researchers are advised to select the most appropriate prompting strategies by the simulation requirements [11, 77, 45, 58]. Upon completion, researchers can obtain a simulated meta-sample, which can then be used to study specific social outcomes.

Step.3 Intelligent Computing Simulation

Intelligent Computing Simulation is the third step that incorporates variables or mechanisms concerning actors, actions, and social environments into the meta-sample. This is done in accordance with theories from the knowledge world and research needs. The simulation compares the results of different settings, analogous to the methodologies employed in experimental and social simulation.

Step.4 Generated Data Analysis

The last step is Generated Data Analysis, which involves a comparison of the results of computational simulations with actual social outcomes in the physical world. The objective is to validate the outcomes against real-world data and to derive theoretical explanations based on the validity of the simulations. Depending on how the generated data is utilized, this step is functionally akin to large-sample quantitative regression analysis or small-sample qualitative process tracing.

3.3.2 Simulation Validation

Despite the widely recognized potential of LLMs and generative AI in social science simulations, no definitive study has yet confirmed that LLM-based agents can precisely simulate human behavior [7]. The ICSM method requires assessing the validity of the research data. To this end, the Simulation Validation path examines the process of generating research data in four steps.

Step.1 Validation Benchmark Setting

The first step is to establish a validation benchmark by identifying a classic research or statistical objective indicators within the knowledge world that examines social outcomes similar to those observed in the ICSM study, using the research's degree of fit as a benchmark to assess the validity of ICSM. Since ICSM is fundamentally a computational simulation method, it is strongly advised that ICSM be employed primarily in analyzing social outcomes studied using classical computational simulation techniques, such as Agent-based modeling (ABM) or cellular automata.

Step.2 Simulation Reliability Test

The second stage is testing the reliability of the simulation. This stage requires running multiple independent trials of the agent's behavior in a well-specified version of the large language model to assess the consistency of the model output. The objective of this process is to ensure the reliability of the simulation.

Step.3 Simulation Validity Assessment

The third step is evaluating the simulation's validity. In this phase, the simulated meta-sample is utilized to conduct computational simulations and assess its goodness of fit to social outcomes compared to the established validity benchmark. ICSM's Reliability Assessment is similar in the evaluation of large language models where new model is accepted based on whether it achieves performance improvement in a specific task [15]. Therefore, we provide two validity test criteria: For experiments using the simulation results of classic studies such as ABM as the baseline, the condition for its effectiveness is to improve the fitting effect of the classic study. For experiments using objective statistical indicators such as election results as the baseline, we recommend that the meta-sample be considered valid only when the fit reaches at least 75%.①

Step.4 Knowledge Discovery Validation

The fourth step is the validation of knowledge discovery. This step establishes the fundamental principle for identifying new variables or mechanisms with theoretical value. The rationale is that a newly introduced variable or mechanism is deemed valid within the knowledge world only if its incorporation enhances the fit beyond the meta-sample.

3.3 Methodological Potentials and Advantages

The ICSM represents the convergence of knowledge, theory, and methods from the social sciences and the application of LLM. The essence of ICSM lies in its capacity to guide LLM's intelligent computing by implementing a scientifically designed research framework while

---

① The 75% baseline for ICSM is derived from three classic experimental results utilizing large models to simulate human actors. The first experiment, conducted at the micro-behavior level, assesses whether LLMs can effectively simulate individual behavior in real social movements to create and support specific ideological content. Across three test datasets comprising real data, the average simulation accuracy of the LLMs reached 72%, systematically surpassing the ABM experimental results (Xinyi Mou, et al, 2024). The second experiment, carried out at the macro-emergence level, employs one million agents to simulate the information propagation path on social media across 198 real-world instances, encompassing the scale, depth, and maximum width of propagation. The experiment calculates the standardized root mean square error (RMSE) between each simulation and the real-world indicator curve, averaging to 75% (Ziyi Yang, et al,2024). The final experiment utilizes multi-dimensional benchmarks to evaluate the capabilities of multiple top commercial large models as agents. This experiment, known as Agentbench, operationalizes the LLMs' ability as intelligent agents to successfully complete reasoning and decision-making tasks in specific situations, thereby providing a generalizable quantitative benchmark for ICSM. Agentbench reveals that GPT-4 achieves a task success rate of 78% in a multi-round open-generation simulation environment(Xiao Liu, et al,2023).By averaging the simulation accuracy of these three classic experiments, we establish a 75% baseline for ICSM.

leveraging AI-agent-generated data to create meaningful simulation data.[①] The fundamental value of ICSM is not merely the application of AI technology but its distinctive epistemic advantage in elucidating social outcomes. In social science, explaining complex social phenomena, including war and peace, the rise and fall of nations, and institutional change, represents a significant challenge. As human society is centered around the evolution of ideas, an explanation of social outcomes must necessarily entail an analysis of the actors involved, the environments in which they operate, and the behavioral mechanisms that underpin their actions. In addressing this problem, a common practice is to simplify social phenomena using various research methods [70], which results in methodological divergence of different epistemological stances. These include quantitative research methods aligned with positivism and qualitative research methods aligned with interpretivism [62]. The former and the big data approach constitute a research paradigm that emphasizes the description of causal effects from a factor-oriented perspective. The latter and the social simulation approach constitute a research paradigm that analyzes causal processes from a mechanism perspective.[②] Although quantitative and qualitative methods are highly complementary, they differ significantly in epistemological stance, research goals, methodological principles, and specific technical tools, which challenge their integration [63]. Quantitative methods are based on numerical data that reflect observations and employ statistics and probability theory for analysis, following deductive logic. In contrast, qualitative methods are founded upon textual data that reflects interpretive concepts. They use the language of logic and set theory for analysis, adhering to inductive logic [28]. This gives rise to the question of paradigm movement at the level of epistemology and theory in current attempts to combine those methods [10]. Before the advent of LLMs, this issue was technically intractable, as researchers either lacked sufficient subjective data or could not compute the aggregated outcomes of these data in a way that closely mimics human cognitive processes.[③] Alternatively, researchers could not compute the aggregated outcomes of these data in a manner that closely mimicked human cognitive processes. As LLMs are the most extensive sources of subjective data and the most accurate simulators of human cognition available in the social sciences, ICSM is not merely a technical gadget embedded with LLMs. Instead, it is a methodological instrument designed to bridge the gap between quantitative and qualitative paradigms, facilitating a unified approach in social science research.

By integrating the four major social science research paradigms—quantitative, qualitative, simulation, and big data—ICSM illustrates the advantages of the fifth paradigm, AI4SS (AI for Social Science). Firstly, at the factor level, ICSM is capable of meeting traditional quantitative

---

① It should be noted that the ICSM method is highly dependent on prompt engineering for few-shot learning. ICSM assumes that scientifically designed prompts can effectively guide LLMs to generate precise responses that reflect human ontological understanding. See related discussion in (Brown et al. 2020).

② Social science research paradigms can be categorized as quantitative, qualitative, social simulation, and big data paradigms. These correspond to the scientific research approaches of experimentation (Empirical Science), theory (Theoretical Science), simulation (Computational Science), and big data (Data-Intensive Science). (Hill 2021)

③ The subjective data includes not only intentional data but also perceptions of the external environment on individual level.

research goals by examining the impact of different factors on agents' behaviors or social causality. Furthermore, the inherent ability of LLMs to process multimodal data can expand the analytical capabilities within the big data paradigm.① Secondly, at the mechanism level, ICSM is designed within the technical framework of social simulation, enabling it to effectively accomplish the goals of qualitative research by identifying and examining mechanisms. As LLM-embedded agents can interact using natural language, ICSM provides the most detailed evidence for mechanism identification and process tracing at the micro level, further enhancing the research advantages of the social simulation paradigm.② Thirdly, ICSM provides a reusable, adjustable, and iterative intelligent model for explainable predictions. ICSM combines machine learning, known for forecasting, with social science theories that explain behavior.③ When successfully implemented in research, ICSM generates simulated agents that can act as substitutes for human participants within intelligent simulation frameworks.. These agents provide a robust theoretical explanation for specific social outcomes. Researchers may utilize these established frameworks to make interpretable predictions about social outcomes based on updated data regarding agents and environments. By employing ICSM, social science researchers can continually benefit from LLM technologies while refining and developing social science theories in accordance with the model fit to real-world outcomes, effectively capturing the evolution of social systems.

## 4 Case Study: Computational Simulation of U.S. Elections

Intelligent computing social simulations remain in their early stages, with several areas still requiring clarification, particularly regarding standardized procedures and operational methods. Key questions remain unexplored: How should ICSM be applied to specific research questions? What constitutes an effective computational simulation workflow? This paper addresses these questions by applying the "Three Worlds" and "Two Cycles" framework to U.S. presidential election simulations, examining specific variables and their effects. We demonstrate ICSM's practical application by detailing the construction of simulated environments and agents, while outlining the essential settings and workflows needed for theoretical exploration.

### 4.1 Step1: Simulation Benchmark Setting and Agent Sample Construction

To meet the research goals and validation requirements, this paper refers to two studies that establish a benchmark for the simulation. "*Forecasting Elections with Agent-Based Modeling: Two Live Experiments*" [26]and "*Out of One, Many: Using Language Models to Simulate Human Samples*" by Argyle et al [4]. The initial study employed Agent-Based Modeling to predict the 2020 U.S. presidential election and the Taiwan region's leadership election, yielding favorable

---

① Big data's core advantage is that it allows for frequent, controlled, and meaningful observations of real-world phenomena using large-scale, multimodal, exhaustive, and fine-grained data (Kitchin 2014; Chang et al. 2014).

② The social simulation paradigm's advantage lies in its ability to capture the emergent mechanisms of meso- and macro-social outcomes within complex social systems while identifying the bidirectional processes linking micro and macro levels (Squazzoni et al. 2014; De Marchi and Page 2014).

③ For the advantages of machine learning in social forecasting and the epistemic value of prediction, see (Hindman 2015; Chen et al. 2021; Wang and Tang 2021; Chen and Liu 2012).

outcomes across six U.S. states.[①] In social sciences, ABM is a method that simulates specific social processes or phenomena by modeling micro-level agents. The modeling process involves setting attributes, defining behavior rules, establishing environmental conditions, and specifying interaction rules. Among these, behavior rules are the core of agent decision-making, determining how agents respond in particular situations. These rules can be simple, condition-based "if… then…" statements, or more complex decision-making processes driven by machine learning or reinforcement learning algorithms[19]. Specifically, Gao et al.'s article employs univariate regression to identify the linear relationships between relevant attributes and outcomes. They define the voting preference ranges for specific agents based on p-values and the signs of correlation coefficients, and then refine the experimental settings by comparing them with historical real-world results. The study constructed state-level samples at a 1:10 ratio (sample size to actual voter population), establishing the identity attributes of the agents based on variables such as age, gender, education, occupation, and race. Subsequently, the voting rules for the agents were defined.

The second study assesses the extent to which LLMs can accurately simulate human samples in terms of their "algorithmic fidelity." The use of American National Election Studies (ANES) data to create synthetic samples of human voters revealed that these samples exhibited opinions and behavioral patterns aligned with those observed in actual survey data. As the two studies share comparable research objectives and methodologies with the present study, their research design and findings are adopted as simulation benchmarks for this study.

Unlike most machine learning tasks where evaluation benchmarks can be obtained through data labeling, establishing accessible and effective metrics for evaluating simulation settings is extremely challenging under the intelligent social science paradigm. However, without an accurate metric, it becomes impossible to evaluate different design schemes before the relevant phenomena occur, making scientific simulation difficult to achieve. The application of ICSM faces the same challenge.

We attempt to address this issue from two aspects: First, we compare our simulation results directly with historical outcomes by measuring the error between predicted and actual results, considering settings with smaller errors to be more accurate and reliable. Building upon this foundation, we compare our work with similar simulation studies. By examining the differences in prediction errors between ICSM and ABM methods, we can establish greater confidence in ICSM's effectiveness when it achieves smaller error margins in simulating real results.

Based on the ABM simulation predictions across states, the average error between this scheme and the actual results is 0.0419. We set this error from the baseline study as our threshold benchmark: only experimental designs with average errors below this value can proceed to the next stage of knowledge exploration.

Once benchmarks have been established, the initial phase of the ICSM experiment entails the

① This study experimented with six states: Michigan, Ohio, Pennsylvania, Indiana, Virginia, and Missouri.

construction of agent samples to create a "simulated society" that accurately reflects the actual demographic distribution. This process requires the identification of key variables and the creation of "background stories" for different agent subgroups to ensure that the overall sample reflects the real-world distribution of these variables. This study identifies the critical variables affecting elections: ethnicity, gender, age, region, education, occupation, and industry. The data utilized in this study is derived from the 2019 American Community Survey (ACS) conducted by the U.S. Census Bureau, which provides an overview of the distributions of these variables across all states.[①] The ACS offers a thorough account of the population sizes of each demographic group across the variables. To ascertain the distribution of variables across states, researchers need only conduct straightforward proportion calculations. It should be noted that while this study uses ABM method as a comparison benchmark, existing ABM research has not included populous states and key swing states due to cost considerations. To verify LLM's simulation capabilities across different types of states, we selected the same six states as existing research for our control experiment: Pennsylvania, Ohio, Michigan, Missouri, Indiana, and West Virginia. For the basic experiment (which only uses identity information to guide agent voting), we additionally evaluated the simulation performance for several large states and swing states including California, Texas, Wisconsin, and Georgia. The construction of agents for all states follows the same logic and approach. The simulation replicated the demographic distribution, with the descriptive statistics for the major variables across the states presented in the Figure. 2.

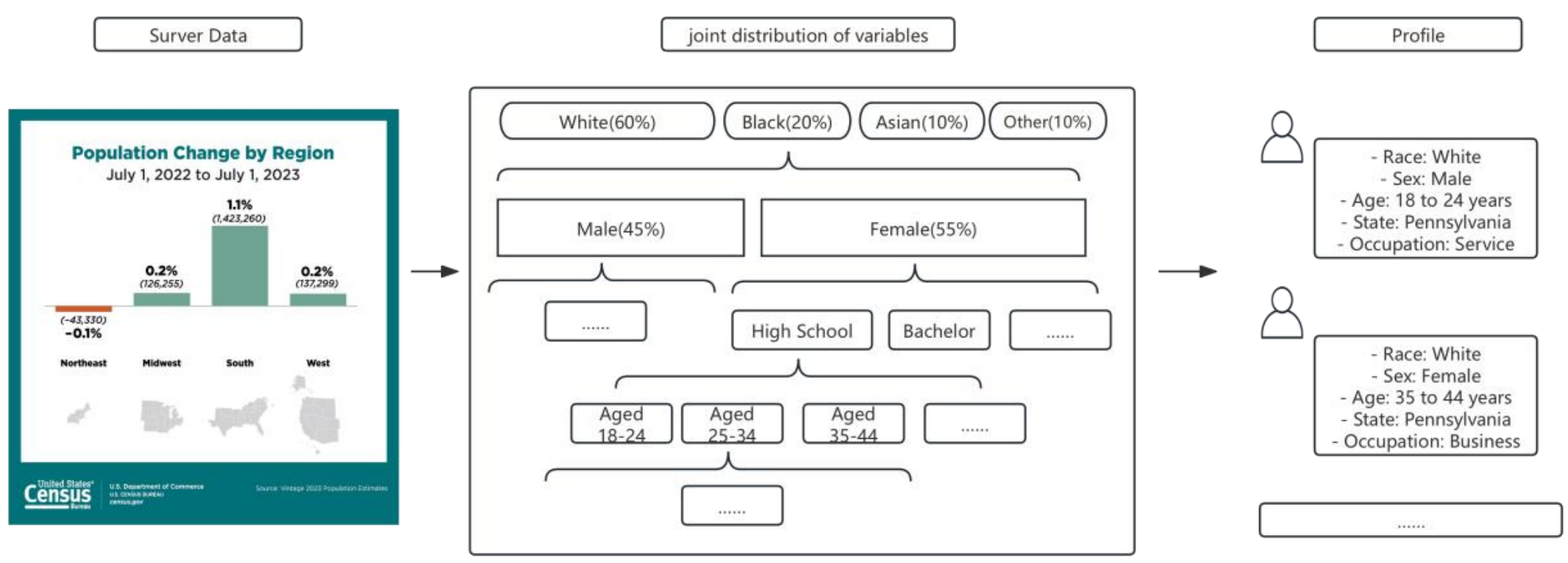

**Fig. 2** The process of generating agent profiles based on real data distribution[②]

---

① The rationale for using the 2019 data is as follows: First, due to the impact of the COVID-19 pandemic, data collection in 2020 was conducted through experimental estimation, resulting in lower data volume and fewer variables than in previous years. Second, ACS data for a specific year is generally released in the fall of the following year (September to December). Hence, even if the 2020 dataset were complete, it would not have been available until the fall of 2021. To boost the adaptability of the experimental design and forecasting potential for future events, the paper opts for the 2019 ACS dataset for constructing the simulated society.

② The elements in the survey data section are taken from the U.S. Census Bureau's official website: US Census Bureau. https://www.census.gov/en.html (Accessed February 14, 2025).

The study employed statistical calculations on all variables with the potential to yield overlapping results, subsequently proportionalizing the cross-variable outcomes across agent samples. This was done with the objective of assigning "background stories" to agents based on the corresponding proportions. This methodology creates "simulated subpopulations with maximum similarity" relative to their real-world counterparts. The generated agent sample is fully aligned with real-world demographic data at the group level and partially aligned at the subgroup level.

From the simulation perspective, the construction of the agent sample is based on clear theoretical foundations and corresponds to real-world data.① The selection of relevant variables is informed by theoretical perspectives and related social science research, with agents' identities assigned based on actual demographic results. The experimental design is consistent with the principles of knowledge and physical reality, satisfying the simulation validation criteria.

**4.2 Step 2: Action Simulation and Simulation Reliability Testing**

Upon completion of the agent sample construction, researchers must conduct action simulation through prompt engineering. Following the methodology proposed by Argyle et al. (2023), we employed zero-shot prompting. We capitalized on LLMs' instruction-following capacity to further delineate AI agents' voting behavior based on identity variables. It is crucial to acknowledge that, due to the vast quantity of training data and the complex parameters of LLMs, it is essential to impose primary constraints on agents' action patterns and information processing to guarantee that the agents retrieve the intended information and execute the precise action as instructed by the researchers. Figure. 3 shows the prompt used in this section.

# Requirement

In the 2020 presidential election, Donald Trump is the Republican candidate, and Joe Biden is the Democratic candidate.

ONLY based on the profile and the information provided above, predict the probability that you would vote for each candidate as well as the probability that you would ”vote for another candidate or not vote at all”. Make sure the probabilities add up to 1.

**Fig. 3** Prompt engineering for agent construction and action setting

As an attention model, LLMs' inferential ability decreases with the addition of irrelevant information, a characteristic similar to "collinearity" in regression models [65]. Related simulation studies also indicate that adding numerous identity features does not improve simulation performance [55]. Therefore, retaining only variables that contribute to prediction is a more appropriate approach. Regarding variable selection, we first referenced existing ABM research and chose similar variable combinations. Additionally, we conducted ablation experiments to

① We selected the Qwen-4-28 large language model to construct agents in electoral scenarios.

screen variables with inconsistent data sources. The relevant experimental results can be found in the Supplementary Materials.

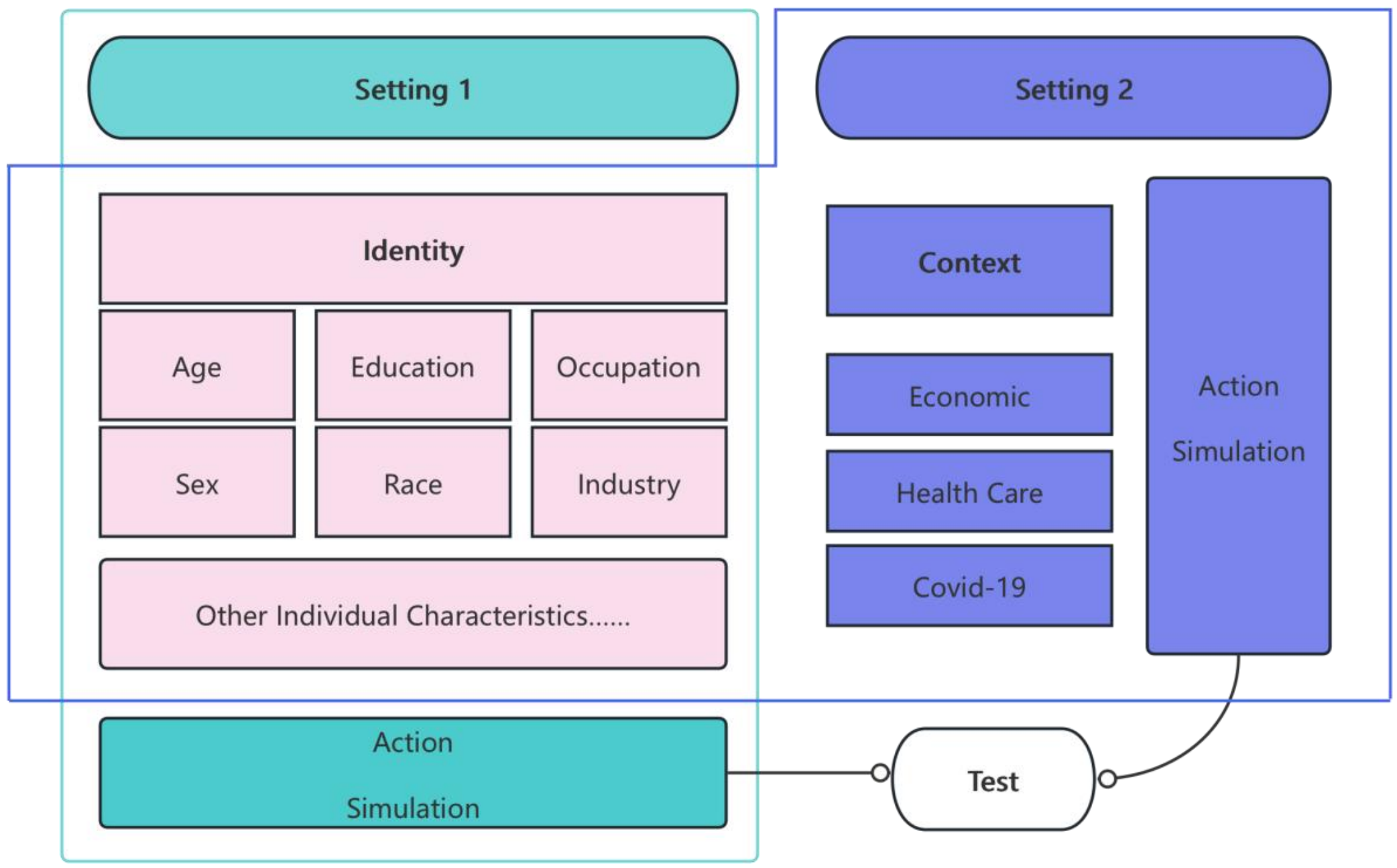

**Fig. 4** Different Experimental Settings

To capture the impact of macro events, we incorporated multiple issues of highest voter concern from national polls in our prompt engineering, combining party platforms, candidate statements, and voter group profiles for simulation settings. Figure. 4 illustrates the two main experimental settings of ICSM, distinguished by whether they include macro-contextual information.

Through analysis of polling data from organizations like Pew Research Center and Gallup, we identified issues of high voter concern: economy, healthcare, and COVID-19 response [5, 25]. Next, based on the 2020 party platforms of both parties, we provided corresponding descriptions for each issue to ensure that each topic reflected the core positions of both candidates and their parties. Finally, to achieve preference activation, we described potential supporter groups related to policy positions. For example, Biden's economic recovery plan appealed to "voters prioritizing equity and wage growth," while Trump's economic plan would "attract voters focused on rapid economic recovery and job creation." Figure. 5 shows the composition of context information.

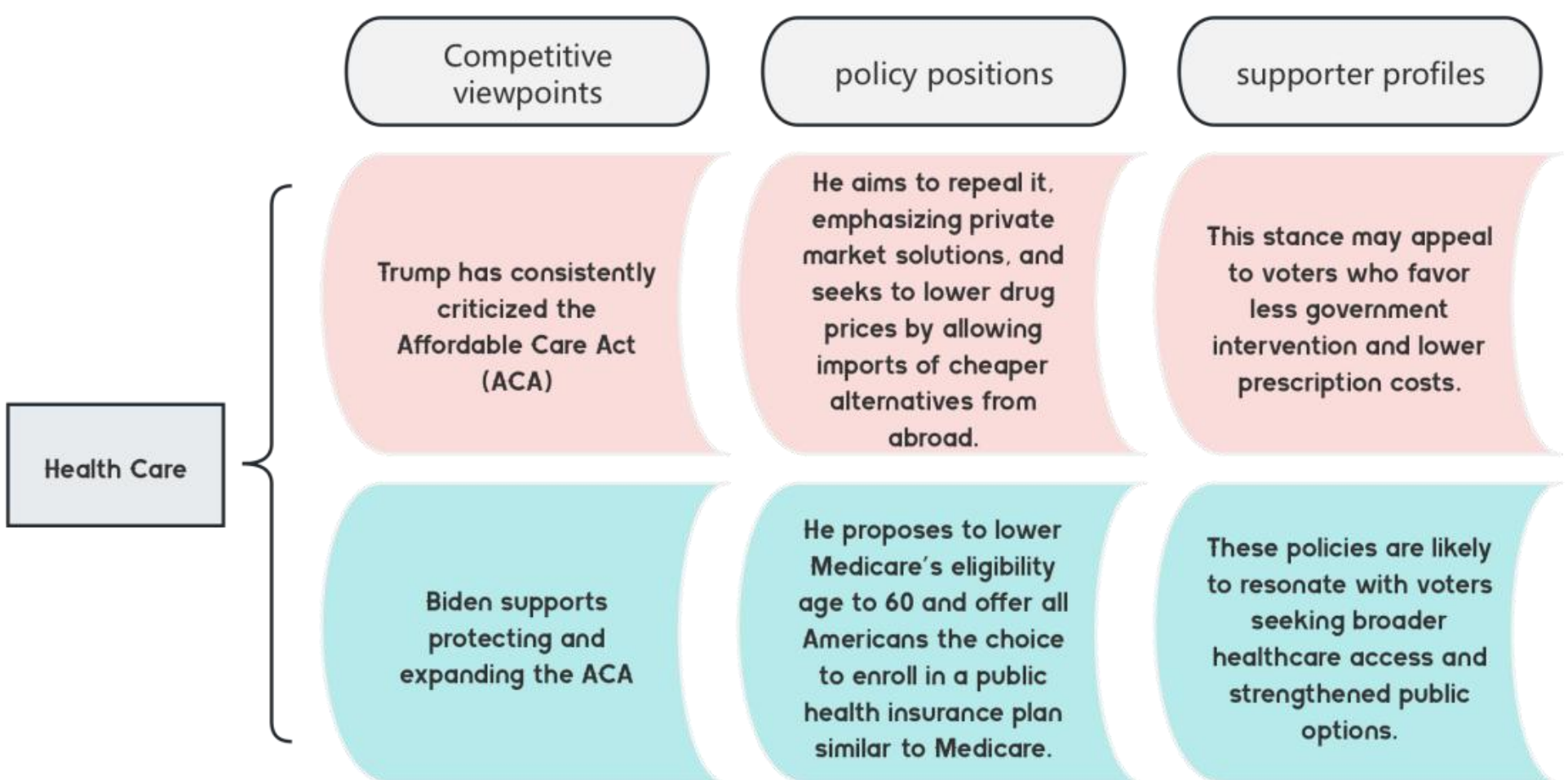

**Fig. 5** Construction method of context information

The reliability of simulation must be evaluated through both action settings and data results. Researchers should provide clear operational definitions of action settings to ensure they produce distinct and definitive results. In election simulations, the process of individual choices aggregating into election results is relatively straightforward, making the connection between behavioral settings and research subjects more intuitive. While this step becomes more complex for abstract concepts, clear operational definitions remain essential to ensure model reproducibility. Building on this foundation, reliability testing requires researchers to document how model results vary when experiments are repeated.

To verify the stability of our results when using large language models, we repeated the profile-only experiment 30 times with 1,000 AI agents per state in four swing states and two populous states.[①] Therefore, we covered states that were difficult to simulate in terms of results and cost considerations. We deliberately set 20-days intervals between repetitions to address concerns about potential variations in model responses across different real-world timestamps. Our main concern is whether experimental results across different rounds, using specific models and prompts, show too much variance to draw reliable conclusions. We used agents' Democratic Party support levels as the main indicator to assess model stability across repeated experiments. The results, including mean and standard deviation of Democratic support, showed minimal variation across state-level results with a maximum fluctuation of 0.002 at a 95% confidence interval.

[①] Different settings actually represent different experiments. When a specific experimental setting is replicated multiple times, each iteration is termed a 'round'.

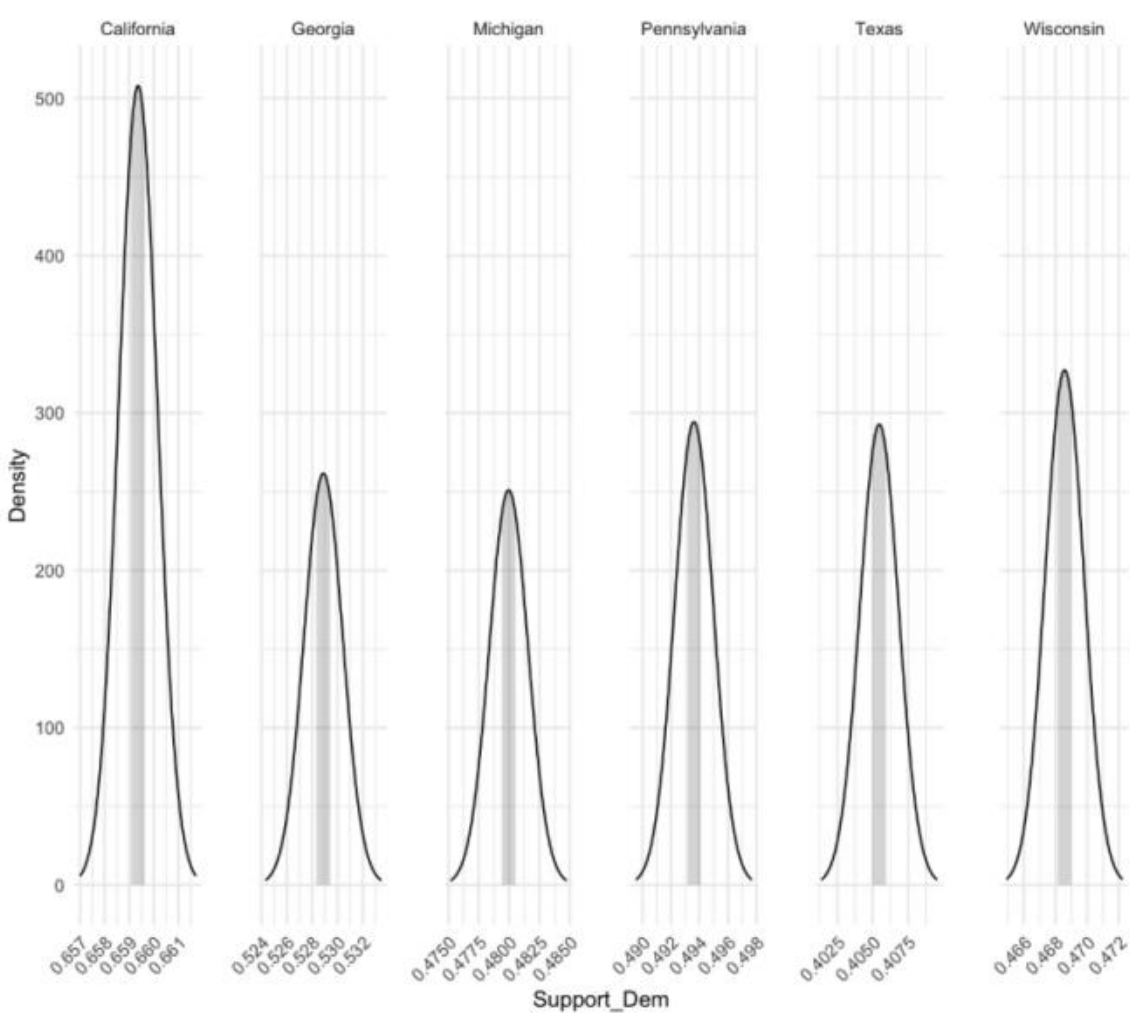

**Fig. 6** Distribution of multi-round experiment results

The sample mean for the state of Michigan, which exhibits the most significant fluctuation, is 0.479. At the 95% confidence level, the confidence interval for Democratic support in the state is [0.480, 0.481]. The model has been shown to exhibit high stability in repeated experiments when prompted with suitable inputs (as shown in Figure. 6), thereby meeting the requirements for reliability testing in simulations.① The data generated by the agent and utilized for analysis in this study fall within the established confidence intervals, which have been subjected to a reliability assessment.

**4.3 Step 3: Intelligent Computing Simulation and Simulation Validity** Assessment

LLM-driven agents can respond to natural language by leveraging agent construction and action simulation, which are subsequently aggregated as computational simulation data within ICSM. The generated data analysis may be conducted regarding different identity variable assignments and action patterns, which may be treated as distinct model settings. By comparing the experimental results with benchmarks (either real-world data or results from referenced studies), ICSM can evaluate the validity of the model configurations. Modifications to the experimental design may entail the introduction of novel identity variables or imposing alternative actions for the agents. A simulation is deemed valid if the experimental results exceed the benchmark. Moreover, if such experimental modifications enhance the simulation's efficacy, the corresponding identity variables and actions can be regarded as theoretically and practically influential factors. Such findings are regarded as constituting "valid knowledge" discovered through experimentation. Through a series of iterative experiments, we evaluate the model's simulation validity by comparing its performance with the results of the ABM study (as illustrated in the table below).②

① This study also evaluated the stability of generated data across different models. In June 2024, we conducted experiments based on GPT-4-Turbo and Qwen-Max models using the same prompt. The average difference in results was 0.007, indicating the stability of the generated data under the same prompts in a certain period.

② All experiment data were normalized to compare experimental results with research benchmarks.

We selected the same states as ABM for the experiments and deployed 300 samples in each state.[①] Table 1 shows the comparison between different experimental settings and benchmarks. The initial experiments (with setting 1, labeled as ICSM_1) constructed agents using only identity characteristics, without providing additional information to influence their reasoning before setting actions. This simulation produced significant errors — with a 0.0727 error rate far exceeding our pre-established baseline — and thus cannot be considered effective. Follow-up experiments (with setting 2, labeled as ICSM_2) enhanced the model by adding relevant policy issues and contextual information. This greatly improved the agents' understanding of the electoral environment and their behavioral reasoning effectiveness. With final error results falling below the ABM baseline, this second approach proved to be an effective simulation setup.

**Table 1** Comparison of ICSM experiment results with the ABM study results

| State | setting | Biden | Trump | Error |
|---|---|---|---|---|
| Ohio | ABM | 0.4925 | 0.5075 | 0.0339 |
| | ICSM_2 | 0.4738 | 0.5262 | 0.0146 |
| | ICSM_1 | 0.5203 | 0.4797 | 0.0610 |
| Missouri | ABM | 0.444 | 0.556 | 0.0234 |
| | ICSM_2 | 0.456 | 0.544 | 0.0358 |
| | ICSM_1 | 0.5126 | 0.4874 | 0.0922 |
| Indiana | ABM | 0.4835 | 0.5165 | 0.0655 |
| | ICSM_2 | 0.4541 | 0.5459 | 0.0357 |
| | ICSM_1 | 0.5050 | 0.4950 | 0.0866 |
| Pennsylvania | ABM | 0.5204 | 0.4796 | 0.0151 |
| | ICSM_2 | 0.5075 | 0.4925 | 0.0016 |
| | ICSM_1 | 0.5386 | 0.4614 | 0.0326 |
| Michigan | ABM | 0.5454 | 0.4546 | 0.0321 |
| | ICSM_2 | 0.485 | 0.515 | 0.0291 |
| | ICSM_1 | 0.5372 | 0.4628 | 0.0231 |
| West Virginia | ABM | 0.3831 | 0.6169 | 0.0812 |

[①] Through our experiments, we found that 300 samples are sufficient to stabilize the experimental results for a state. Related content can be found in the supplementary materials.

| | ICSM_2 | 0.3845 | 0.6155 | 0.0824 |
|---|---|---|---|---|
| | ICSM_1 | 0.5568 | 0.4432 | 0.1409 |
| Mear error | ABM:0.0419 | | ICSM_1:0.0727 | ICSM_2:0.0332 |

### 4.4 Step 4: Generated Data Analysis and Knowledge Discovery Validation

Output results mark the endpoint of traditional prediction paths, but not for ICSM. In most cases, prediction work is data-driven and therefore "de-theorized," where the "learning" process discovers data patterns rather than theoretical patterns. These data patterns cannot be theoretically explained and need to be updated as data changes. In contrast, ICSM places theory at the core of simulation prediction. Effective knowledge is that which achieves simulation improvements in specific contexts. When we employ rigorous, evidence-based benchmark settings, establish clear action parameters, achieve repeatable and stable experimental results, and demonstrate improved error rates compared to existing research, we feel safe to classify these samples as “effective” with potential for further knowledge exploration. Built on agent instruction-following characteristics and well-defined experimental paths, ICSM can be applied across diverse scenarios and integrated with various methodological approaches. In this electoral scenario study, we first examine the use of generated voter agents for quantitative analysis in counterfactual frameworks. Then, we focus on the possibility of obtaining fine-grained micro-evidence by these agents to support mechanism analysis for middle-range theories. Finally, we emphasize that this useful analytical framework can be applied to other states in a relatively straightforward yet effective manner.

(1) Quantification: Counterfactual Experiments Based on ICSM

While causal inference drives quantitative research, it faces a fundamental paradox — counterfactuals don't exist in reality. Following the principle "no causation without manipulation," randomization and intervention offer the most reliable path to causal inference. Random assignment creates groups with matching characteristics, allowing us to attribute outcome differences to the intervention itself. When individual counterfactuals aren't possible, comparing control and treatment groups at the group level offers a practical alternative. Though a real person can't be in both control and treatment groups simultaneously, an artificial agent can.

In our simulation experiments, we maintain consistent agent behavior through independent API calls and identical identity and context parameters.[1] Our intervention design allows agents to experience different external conditions "simultaneously."

Our theoretical exploration examines elements that reduced mean absolute error in simulation design. In the context of improving model predictions through economic issues, we first investigate voters' actual attitudes toward taxation. While tax policy is closely tied to America's growing social inequality, voters' views on taxation may differ from their stance on social equality due to the politically sensitive nature of social justice.[①] We therefore ask: How much did liberal tax policies shape political preferences in the presidential election? We explore this through a counterfactual framework. We employ list experiments with random assignment and list intervention to reveal group-level tax preferences.

[1] Independent calls can be viewed as "counterfactuals" since the agent's behavior and choices remain stable under identical settings. To verify this stability, we called the same agent 100 times independently and analyzed the support results. The outputs showed a mean support rate for Biden of 0.533, with a standard deviation of 0.0457 and a range of 0.524-0.542 within a 95% confidence interval. These consistent results support our assertion that independent calls can effectively construct counterfactuals for specific samples.

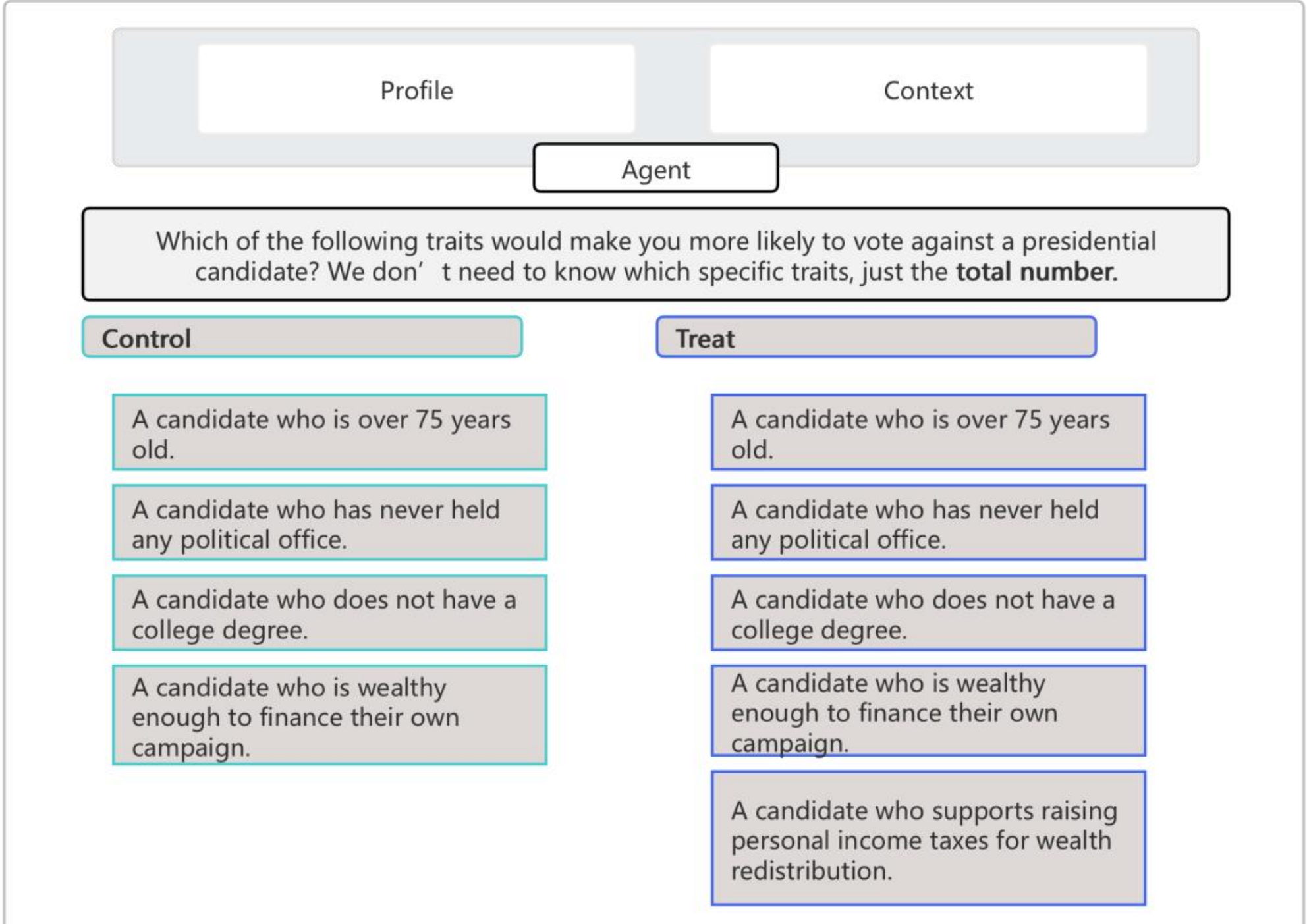

**Fig. 7** Construction of List Experiments

Figure. 7 illustrates the operational logic of the LIST experiment. As mentioned above, we construct agents through context and profile. After this, through independent API calls, we enable the same agent to simultaneously enter both control and treatment groups. We investigate which characteristics influence an agent's support for a presidential candidate. The key setup is that for the treatment group, critical questions are added on top of all the control group questions, so that based on random assignment, the difference between the two groups represents the true group attitude.Our basic experimental design builds on research by [13], while addressing the core question: "A candidate who supports raising personal income taxes for wealth redistribution.".

Each agent receives assignments to both control and treatment groups through separate calls, reporting their number of concerning issues in each instance. Our experiment used 300 samples from Pennsylvania, yielding 600 total observations.

[①] Kochhar, Juliana Menasce Horowitz, Ruth Igielnik and Rakesh. 2020. Most Americans Say There Is Too Much Economic Inequality in the U.S., but Fewer Than Half Call It a Top Priority. Pew Research Center. https://www.pewresearch.org/social-trends/2020/01/09/most-americans-say-there-is-too-much-economic-inequality-in-the-u-s-but-fewer-than-half-call-it-a-top-priority/ (Accessed February 11, 2025).

**Table 2** Regression Results of List Experiments

| | Dependent variable: count | Standard Error |
|---|---|---|
| Treat | 0.787*** | 0.050 |
| Race | 0.167** | 0.067 |
| Sex | 0.018 | 0.050 |
| College | 0.054 | 0.051 |
| Age | -0.136*** | 0.013 |
| Constant | 4.219*** | 0.094 |
| Observations | 600 | |

Note: *p<0.1; **p<0.05; ***p<0.01

The experimental results show that the control group's average response value was 3.8, while the treatment group's was 4.6. In Table 2, the supplementary regression model shows robust and significant results, suggesting that voters' attitudes toward taxation in the 2020 election may have been more conservative than conventionally understood. Real polling data provided similar observations, indicating relatively conservative tax attitudes among voter groups.① We further examine whether the patterns in our generated data correspond to real-world data.

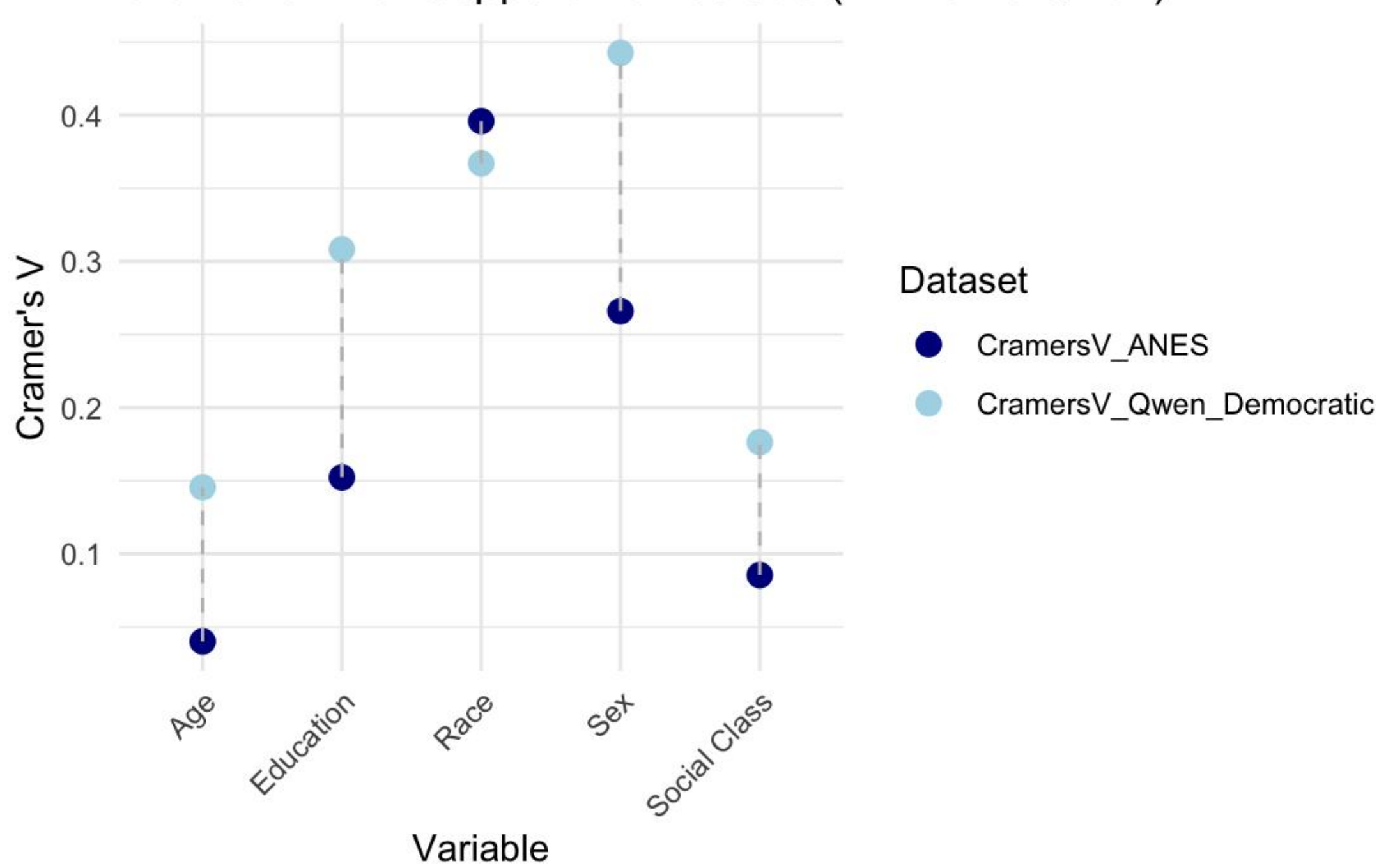

**Fig. 8** Comparison of Cramer's V correlations for "supporting the Democratic Party" between ANES and Qwen data

① ITR. 2020. Trump or Biden? 66% Say Taxes Are a Deciding Factor. Iowans for Tax Relief. https://taxrelief.org/trump-biden-taxes/ (Accessed February 11, 2025).

The most straightforward method for evaluating ICSM-generated data is to conduct a comparative analysis with real-world datasets. This study employs the long-standing, high-quality, and widely utilized survey research project, the American National Election Studies (ANES), as a benchmark [39, 23]to investigate the relationship between identity variables and support for the Democratic Party in both the ICSM-generated and actual data.[①] The "Cramer's V" coefficient was selected as the indicator, as it is suitable for examining the correlation between categorical variables.[②] As a consequence of the discrepancies in the definition of the dependent variable and the method of variable encoding, it is not anticipated that the coefficients will correspond fully. Nevertheless, the analysis demonstrates a general correspondence in the overall patterns between the two datasets. As shown in Figure. 8, the stability of Cramer's V coefficients across variables indicates that even without ICSM-generated data, researchers can leverage this data for exploratory analysis, providing valuable support alongside traditional surveys or big data text analyses.

(2) Qualification: Mechanism Identification under ICSM

Beyond quantitative approaches, by fully utilizing the agents' instruction following capabilities, we can also obtain qualitative mechanism information. In election research, mechanism identification relies on micro-level information provided through questionnaires and interviews. For a sample of agents that effectively simulated the 2020 election results, we conducted a complete survey simulation. The questionnaire design is shown in the Figure. 9, covering 100 agent samples from Pennsylvania.

---

① Based on the ACS and ANES data structures, the independent variables in this study are treated as categorical. This variable covers all educational levels.

② Agent samples in this study's experiments were given a binary choice: support the Democratic or Republican Party. If an agent doesn't support the Democratic Party, it’s counted as supporting the Republican Party, and vice versa. Therefore, showing only the "supporting the Democratic Party" results in the data analysis still represents the overall experimental outcomes.

We would like to conduct a survey regarding your voting behavior in the 2020 Presidential Election. Your responses will help us better understand voting patterns and preferences across different socio-economic groups. Please answer the following questions as accurately as possible.

# Section 1: Basic information

What is your racial/ethnic background?

Have you received higher education?

How would you describe your social class? Are you in the lower class, the working class, the middle class, or the upper class?

# Section 2: Alignment

Please say to what extent you agree or disagree with the following statement:
'The government should take measures to reduce differences in income levels'.

Do you agree strongly, agree somewhat, neither agree nor disagree, disagree somewhat, or disagree strongly?

Does the increasing number of people of many different races and ethnic groups in the United States make this country a better place to live, a worse place to live, or does it make no difference?

# Section 3: Vote

Considering your educational background, please indicate which candidate’s policies you prefer for each policy issue and explain why.

Additionally, summarize your reasoning with three keywords for each issue.

**Fig. 9** Construction of Survey Questionnaire

Understanding mechanism as the process leading from cause to effect, we focus on how educational background influences voters' policy preferences and thereby their political stance. Through questions in Part 3 of the survey, we obtain information about voters' preferences for different policies and their related explanations. By organizing keywords, we can gain a comprehensive understanding of the mechanism pathways. Combined with keywords and response content, we can develop a deeper understanding of these mechanism pathways.

The Cramer coefficient mentioned above has shown that education level positively correlates with Democratic Party support, and the survey results statistics further verify this finding. Across all policies, voters with higher education show stronger support for Biden than those without higher education. However, analysis of the response texts reveals more specific information. First, voter groups with different education levels have varying reasons for supporting Biden. Voters with higher education more frequently use keywords like "equity" (89% of agents mentioned) and "fair"(47% of agents mentioned) to describe their reasons for support, with typical responses shown in Figure. 10.

"As a college-educated Black woman working in public administration, Biden's focus on equity, wage growth, and public service investments aligns more with the values of supporting communities and stable job markets."

**Fig. 10** Example Response from College-Educated Biden Supporters

Voter groups tend to evaluate Biden's economic policies more from the perspective of social values and fairness. However, for voters without higher education, support for Biden stems from more explicit economic reasons. "Minimum wage" (89% of agents mentioned) is the most frequently mentioned keyword among this group (as shown in Figure. 11).

"As a high school graduate working in the service sector, stability and accessible opportunities for wage growth are crucial. Biden's proposal to raise the minimum wage and invest in public services could lead to more secure employment conditions and better pay."

**Fig. 11** Example Response from Non-College-Educated Biden Supporters

In conclusion, we identified distinct mechanistic pathways within the same theoretical framework through analyzing keywords and representative responses from both groups. Within the linear relationship between education level and Democratic Party support, we found two distinct narratives that cannot be quantitatively measured: value-based positions and practical interest considerations. This evidence demonstrates that our comparative testing model can contribute to deeper theoretical analysis by conducting qualitative research approach.

(3) Expanding ICSM to Other States

Beyond its contributions to traditional research methods, a key advantage of ICSM is that it can be implemented efficiently and cost-effectively across different states. Even when researchers are interested in states outside those examined in this study, they can adapt our framework for their investigations. Specifically, ICSM's lightweight simulation framework and research design effectively addresses populous and swing states that traditional methods struggle to analyze. While conventional approaches face prohibitive computational costs when simulating new states, ICSM requires only minimal API call costs. This is because ICSM establishes identity-based decision patterns through preliminary training, using data solely to characterize specific agent identities rather than requiring extensive training for each state. Once a model setting meets benchmark standards and demonstrates improved performance, researchers can uniformly apply the setting to simulate results across different states. This method offers clear practical advantages: enabling cost-effective and efficient simulations across multiple states.

Using consistent methods to access population information, we conducted simulations (300 agents per state) in populous states (California, Texas) and swing states (Wisconsin and Georgia), with the results shown in Table 3.

**Table 3** Application of ICSM settings in swing states and populous states

| State | Vote Share | ICSM_1 Error | ICSM_2 Error |
|---|---|---|---|
| Georgia | 0.5012:0.4988 | 0.0700 | 0.0025 |
| Wisconsin | 0.5032:0.4968 | 0.0277 | 0.0245 |
| California | 0.6490:0.3510 | 0.0238 | 0.0848 |
| Texas | 0.4716:0.5284 | 0.0223 | 0.0263 |
| MAE | | 0.0360 | 0.0345 |

The experimental errors across these states fell below ABM benchmarks, validating our design improvements and demonstrating ICSM's high external validity. Through comparing different experimental settings, we found that incorporating education and context information consistently reduced mean absolute error, prompting deeper analysis of these elements. Such detailed investigation remains beyond traditional methods' reach — in the big data paradigm, variables that enhance model performance rarely contribute to knowledge discovery beyond prediction.

## 5 Conclusion: Towards Intelligent Social Science Research Methods

LLMs mark the advent of artificial general intelligence, presenting revolutionary opportunities and methodological challenges for political science. As the AI4SS (AI for Social Science) paradigm gains academic consensus, this paper examines the rationale behind the innovative impact of LLMs on research methods from the perspectives of mechanism, methodology, and application. This paper puts forward a novel approach, Intelligent Computing Social Modelling, driven by LLM. Theoretically, this study enhances our comprehension of how LLM can improve knowledge production in the social sciences. Practically, it introduces an empirically validated, AI-based method for explaining and forecasting political outcomes.

This paper explores the role of LLMs from the perspectives of ontology, epistemology, and social sciences methodology, integrating political science and computer science viewpoints throughout the passage. Firstly, from the standpoint of how technological advancement facilitates the evolution of research paradigms within the social sciences, it emphasizes the value of LLMs in driving methodological innovation. By integrating multimodal data and computational simulations based directly on natural language, LLMs can better support the inductive and deductive methods in achieving their research goals and provide technically robust tools for synthesizing these two logics. Secondly, the paper examines the innovative mechanisms of LLMs from the perspective of idea synthesis and action simulation. LLMs, with their ultra-large-scale training datasets and ability to integrate multimodal data (text, images, speech, and video), can improve opinion collection and analysis tasks traditionally handled by surveys. This is accomplished by developing intelligent agent samples that more accurately reflect human perceptions in the real world. With regard to action simulation, ICL and CoT reasoning techniques permit LLMs to operate as implicit computational models of human behavior and as flexible social context framers. With reasoning abilities approaching those of humans, they can simulate a range of actions in more complex and realistic environments, thereby overcoming the limitations of traditional action research methods, such as formal models and experiments, constrained by their inherent simplicity.

This paper introduces a new methodology, Intelligent Computing Social Modeling, which leverages the advantages of LLMs while merging classic social science research paradigms. This methodology is based on the elucidation of functional mechanisms. The fundamental aspect of this methodology is the construction of an LLM-based simulated society that exhibits a high degree of correspondence with the real world in terms of demographic distribution, social environment, and behavioral patterns. Subsequently, the data generated from the simulated agent sample is employed to investigate particular social outcomes. In accordance with Popper's "three worlds" theory, this methodology is divided into two distinct parts. The two key stages are "simulated society construction" and "simulation validation." The former is completed following the four steps: AI agent sample construction, action simulation, intelligent computing simulation, and generated data analysis. To guarantee the veracity of the generated data, it is essential to conduct a preliminary validation of the simulation. This entails establishing benchmarks for validity, testing the simulation's reliability, assessing the simulation's validity, and validating the knowledge discovered through the simulation. This method enhances the benefits of the quantitative paradigm in applying big data to calculate the impact of factors while simultaneously providing qualitative research with individual-level evidence for social mechanisms. As a result, it offers a methodological tool for developing the fifth paradigm in social sciences. The ICSM method, as evidenced by the empirical case of the 2020 U.S. election, attains greater simulation accuracy than the classic ABM method and facilitates the integration and development of quantitative and qualitative traditions by providing explanatory weights, agent samples, and evidence of individual-level opinion.

The essence of ICSM lies in constructing an agent capable of perceiving specific spatiotemporal context information and emulating the role of a real human actor within the corresponding social system. ICSM represents an intelligent social science research methodology founded on intelligent agents powered by LLMs. The knowledge discovery logic of this approach involves leveraging the correspondences among large models technology, the material world, the mental world, and the knowledge world to investigate social outcomes, actions, and ideas by recreating simulated societies that generate diverse social phenomena. Originating from explorations in political science, ICSM can be extended to at least two pivotal social science tasks. Firstly, it offers explainable predictions. By simulating various types of human actors, ICSM furnishes an explainable prediction tool for election results, policy-making, and public opinion shifts. For instance, by transforming voter simulations into legislator simulations, the ICSM framework can be adapted for predicting policy legislation outcomes. Numerous experiments have substantiated the efficacy of this approach in enhancing prediction accuracy [46, 47]. Building on this, researchers can glean evidence of individual-level behaviors and concepts by further examining and testing agent samples, thereby significantly bolstering the interpretability of predictions grounded in artificial intelligence. Secondly, ICSM provides research samples with high abundance. Given that the simulation trajectory of ICSM entails constructing a cohort of human samples within a specific social context, this group encompasses not only identity variables derived from real statistical data but also a wealth of variable relationships. For these samples, ICSM presents multiple potential applications. When integrated with quantitative methods, ICSM constructs robust counterfactuals through its ability to generate independent and stable samples. In combination with qualitative research, ICSM facilitates the exploration of diverse narratives and attitudes using natural language. As demonstrated by the list experiments, data comparative analysis, and questionnaire surveys, AI voters in election simulations not only display their political preferences but also show clear understanding of external economic policies and their personal economic interests. It is conceivable that through this cohort of voter samples, we can also probe economic variables that are markedly influenced by political preferences.

In summary, ICSM not only facilitates the integration of artificial intelligence science across with social science but also fosters the convergence of political science, sociology, and economics within the social science domain. However, this study is subject to certain limitations. As a simulation-based study, our primary concern is the effectiveness of the simulation itself. However, during the process of generating AI agents, we also recognized that, due to the nature of their training, LLMs inevitably embed certain cultural or political biases [88]. In fact, it is likely that no model can achieve complete neutrality at the ideological level. While we recognize that such biases pose challenges for the safe development of AI, they also, to some extent, preserve certain characteristics of real-world societies within simulations [4]. Whether this embedded bias hinders or enhances the validity of social simulations remains an open question and warrants further exploration. To assess political bias in our experiment, we conducted two versions of a manipulation-based test, where agents were exposed exclusively to the viewpoints of a single candidate. The results revealed that while the two manipulations produced opposite directional effects, the magnitude of these effects remained similar. This suggests that political bias in the Language model applied in this study was not rigid enough to meaningfully distort the outcomes. The detailed results are available in the supplementary materials, and our approach may serve as a valuable reference for future studies. Secondly, the premise of ICSM is that there exists a correspondence between the cognitive patterns and mental models of large models and the real world. Although efforts are made to establish benchmarks based on analogous knowledge and classical human theories at each pivotal stage of large model data generation, this approach necessitates more rigorous testing concerning logical self-consistency and experimental validation.

To enhance the stability and reproducibility of research utilizing large language models (LLMs), several key recommendations should be considered. First, it is essential to ensure a standardized approach when calling LLMs via APIs and to explicitly specify the exact model version being used. Given the continuous updates and iterations of LLMs, failing to document the specific model version can lead to inconsistencies in results across different studies. Second, researchers should design well-defined benchmarks to evaluate outcomes effectively. When working with open-ended text generation, assessing differences across iterations can be inherently challenging. However, anchoring evaluations on quantifiable indicators, such as numerical metrics or proportional changes, can enhance the interpretability and comparability of results. While many research questions cannot be fully reduced to a single numerical measure, ensuring precise definitions of both the results and the simulated subjects can significantly mitigate this challenge. Finally, conducting multi-faceted validation tests is crucial for assessing result stability. Suggested approaches include repeated executions of the same prompt, multi-round evaluations of overall results, and tests conducted over different time periods. The key principle is that researchers should strive for the most comprehensive testing possible to report the robustness of their findings. Looking ahead, we anticipate greater involvement of social science researchers in shaping the development, design, and application of LLMs. Such collaboration will facilitate a deeper integration between LLMs and social science research, ultimately enhancing their methodological synergy.

Future research endeavors can delve deeper into the performance of agents in experiments pertaining to diverse social outcomes, such as policy-making, consumption forecasting, and communication forecasting, to achieve a more comprehensive understanding of the relationship between generated data and observed data. It is projected that by 2025, 10% of global data will be AI-generated. Given the novelty and complexity of LLMs, a clear comprehension of the core concepts, mathematical logic, and cognitive significance of AI-generated data is imperative for advancing high-quality innovation in social science methodology during the intelligent era.

**References**

[1] Abdelnabi, S., A. Gomaa, S. Sivaprasad, L. Schönherr, and M. Fritz. 2023. LLM-deliberation: Evaluating LLMs with interactive multi-agent negotiation games. *arXiv preprint arXiv:2309.17234*.

[2] Agadjanian, A., and D. Lacy. 2021. Changing votes, changing identities? Racial fluidity and vote switching in the 2012–2016 US Presidential Elections. *Public Opinion Quarterly* 85(3): 737-752.

[3] Ahearn, C. E., J. E. Brand, and X. Zhou. 2023. How, and For Whom, Does Higher Education Increase Voting? *Research in Higher Education* 64 (4): 574–597. DOI: 10.1007/s11162-022-09717-4

[4] Argyle, L. P., E. C. Busby, N. Fulda, J. R. Gubler, C. Rytting, and D. Wingate. 2023. Out of One, Many: Using Language Models to Simulate Human Samples. *Political Analysis* 31 (3): 337–351. DOI: 10.1017/pan.2023.2

[5] Atske, Sara. 2020. 4. Important issues in the 2020 election. Pew Research Center. https://www.pewresearch.org/politics/2020/08/13/important-issues-in-the-2020-election/. Accessed 3 February 2025.

[6] Austen-Smith, D. and J. S. Banks. 1998. Social choice theory, game theory, and positive political theory. *Annual Review of Political Science* 1 (1): 259-287. DOI: 10.1146/annurev.polisci.1.1.259

[7] Bail, C. A. 2024. Can Generative AI improve social science? *Proceedings of the National Academy of Sciences* 121 (21): e2314021121. https://doi.org/10.1073/pnas.2314021121

[8] Bendor, J. 2003. Herbert A. Simon : Political Scientist. *Annual Review of Political Science* 6 (1): 433–471. DOI: 10.1146/annurev.polisci.6.121901.085659

[9] Bendor, J., D. Diermeier, and M. Ting. 2003. A behavioral model of turnout. *American Political Science Review* 97 (2): 261–280;

[10] Brannen, J. 2016. Combining qualitative and quantitative approaches: an overview in *Mixing Methods: Qualitative and Quantitative Research*, ed. Brannen, J., 3-37. London: Routledge.

[11] Brown, T. B., B. Mann, N. Ryder, M. Subbiah, J. Kaplan, P. Dhariwal, A. Neelakantan, P. Shyam, G. Sastry, A. Askell, S. Agarwal, A. Herbert-Voss, G. Krueger, T. Henighan, R. Child, A. Ramesh, D. M. Ziegler, J. Wu, C. Winter, M. Chen, E. Sigler, M. Litwin, S. Gray, B. Chess, J. Clark, C. Berner, S. McCandlish, A. Radford, I. Sutskever, and D. Amodei. 2020. Language models are few-shot learners. *Advances in Neural Information Processing Systems* 33: 1877-1901.

[12] Campbell, J. E. 2004. Introduction—The 2004 presidential election forecasts. *PS:Political*

*Science & Politics* 37 (4): 733–735. DOI: 10.1017/S1049096504045020

[13] Carmines, Edward G., and Eric R. Schmidt. 2021. Prejudice and tolerance in US presidential politics: Evidence from eight list experiments in 2008 and 2012. *PS: Political Science & Politics* 54(1): 24–32.

[14] Chang, R. M.; Kauffman, R. J.; Kwon, Y. 2014. Understanding the paradigm shift to computational social science in the presence of big data. *Decision Support Systems* 63: 67–80. DOI: 10.1016/j.dss.2013.08.008

[15] Chang, Y., X. Wang, J. Wang, Y. Wu, L. Yang, K. Zhu, X. Yi, C. Wang, Y. Wang, W. Ye, Y. Zhang, Y. Chang, P. Yu, Q. Yang, and X. Xie. 2024. A Survey on Evaluation of Large Language Models. *ACM Transactions on Intelligent Systems and Technology* 15 (3): 1-45.

[16] Chen, D. and F. Liu. 2012. Taking Prediction Seriously: International Relations Theory Development and Theoretical Prediction (认真对待预测:国际关系理论发展与预测). *World Economics and Politics* (01):19-33+155-156.

[17] Chen, Y., X. Wu, A. Hu, G. He, and G. Ju. 2020. Social Prediction: A New Research Paradigm Based on Machine Learning (社会预测:基于机器学习的研究新范式). *Sociological Studies* 35 (03): 94-117+244. DOI: 10.19934/j.cnki.shxyj.2020.03.005.

[18] Coleman, J. S. 1990. *Foundations of Social Theory.* Harvard University Press.

[19] De Marchi, S. and S. E. Page. 2014. Agent-Based Models. *Annual Review of Political Science* 17 (1): 1–20. DOI: 10.1146/annurev-polisci-080812-191558

[20] Dillion, D., N. Tandon, Y. Gu, and K. Gray. 2023. Can AI language models replace human participants? *Trends in Cognitive Sciences* 27 (7): 597–600. DOI: 10.1016/j.tics.2023.04.008

[21] Dong, Q. and W. Liu. 2023. Cross-Disciplinary Research and Paradigm Shifts: Towards a New Era of AI-Based Analytics in International Relations Studies (学科交叉与范式革新:迈向国际关系研究的智能分析时代). *Chinese Journal of European Studies* 41 (5): 146-162+176.

[22] Easton, D. 1969. The new revolution in political science. *American Political Science Review* 63 (4): 1051-1061. DOI:10.2307/1955071

[23] Erikson, R. 2002. National election studies and macro analysis. *Electoral Studies* 21: 269-281. DOI: 10.1016/S0261-3794(01)00020-8.

[24] Fan, C., J. Chen, Y. Jin, and H. He. 2024. Can large language models serve as rational players in game theory? A systematic analysis. *Proceedings of the AAAI Conference on Artificial Intelligence* 38 (16). DOI: 10.1609/aaai.v38i16.29751

[25] Gallup. 2020. Several issues tie as most important in 2020 election. https://news.gallup.com/poll/276932/several-issues-tie-important-2020-election.aspx. Accessed 3 February 2025.

[26] Gao, M., Z. Wang, K. Wang, C. Liu, and S. Tang. 2022. Forecasting elections with agent-based modeling: Two live experiments. *PLoS One* 17 (6): e0270194. DOI: 10.1371/journal.pone.0270194.

[27] Ghaffarzadegan, N., A. Majumdar, R. Williams, and N. Hosseinichimeh. 2023. Generative

agent-based modeling: Unveiling social system dynamics through coupling mechanistic models with generative artificial intelligence. *arXiv preprint arXiv:2309.11456*. DOI: 10.48550/arXiv.2309.11456

[28] Goertz, G. and J. Mahoney. 2012. *A Tale of Two Cultures: Qualitative and Quantitative Research in the Social Sciences*. New Jersey: Princeton University Press.

[29] Grossmann, I., M. Feinberg, D. C. Parker, N. A. Christakis, P. E. Tetlock, and W. A. Cunningham. 2023. AI and the transformation of social science research. *Science* 380 (6650): 1108-1109. DOI:10.1126/science.adi1778

[30] Hadi, M. U., Q. Al Tashi, R. Qureshi, A. Shah, A. Muneer, M. Irfan, A. Zafar, M. B. Shaikh, N. Akhtar, S. Z. Hassan, M. Shoman, J. Wu, S. Mirjalili, and M. Shah. 2023. A Survey on Large Language Models: Applications, Challenges, Limitations, and Practical Usage. Authorea Preprints.

[31] Hao, S., Y. Gu, H. Ma, J. J. Hong, Z. Wang, D. Z. Wang, and Z. Hu. 2023. Reasoning with Language Model is Planning with World Model. *arXiv preprint arXiv:2305.14992*. DOI: 10.48550/arXiv.2305.14992

[32] Harshvardhan, G. M., M. K. Gourisaria, M. Pandey, and S. S. Rautaray. 2020. A comprehensive survey and analysis of generative models in machine learning. *Computer Science Review* 38: 100285. DOI: 10.1016/j.cosrev.2020.100285

[33] Hill, C. A. 2020. Moving Social Science into the Fourth Paradigm: The Data Life Cycle, in *Big Data Meets Survey Science: A Collection of Innovative Methods*. ed. Hill, C. A., Biemer, P. P., Buskirk, T. D., Japec, L., Kirchner, A., Kolenikov, and S., Lyberg, L. E., 713-731. New Jersey: John Wiley & Sons.

[34] Hindman, M. 2015. Building Better Models: Prediction, Replication, and Machine Learning in the Social Sciences. *The ANNALS of the American Academy of Political and Social Science* 659 (1): 48–62. DOI: 10.1177/0002716215570279.

[35] Horton, J. J. 2023. Large Language Models as Simulated Economic Agents: What Can We Learn from Homo Silicus? National Bureau of Economic Research. Working Paper 31122. DOI: 10.3386/w31122

[36] Hu, A. and S. Zhou. 2024. A Novice Standing on the Shoulders of Giants: Generative AI in Social Science Research (站在巨人肩膀上的初学者:社会科学研究中的生成式人工智能). *Jiangsu Social Sciences* (1): 57-65+242. DOI:10.13858/j.cnki.cn32-1312/c.20240201.001.

[37] Hua, W., L. Fan, L. Li, K. Mei, J. Ji, Y. Ge, L. Hemphill, and Y. Zhang. 2023. War and Peace (WarAgent): Large Language Model-based Multi-Agent Simulation of World Wars. *arXiv preprint arXiv:2311.17227*. https://doi.org/10.48550/arXiv.2311.17227

[38] Huang, J., E. J. Li, M. H. Lam, T. Liang, Y. Yuan, W. Jiao, X. Wang, Z. Tu, and M. R. Lyu. 2024. How Far Are We on the Decision-Making of LLMs? Evaluating LLMs' Gaming Ability in Multi-Agent Environments. *arXiv preprint arXiv:2403.11807*. DOI: 10.48550/arXiv.2403.11807

[39] Jackman, S., and Spahn, B. 2019. Why Does the American National Election Study Overestimate Voter Turnout? *Political Analysis* 27 (2): 193–207. DOI:10.1017/pan.2018.36.

[40] Jarrett, D. and M. Pislar, A. Tacchetti, M. A. Bakker, H. Tessler, R. Koster, J. Balaguer, R. Elie, C. Summerfield and A. Tacchetti. 2023.Language agents as digital representatives in collective decision-making. *NeurIPS 2023 Foundation Models for Decision Making Workshop*.

[41] Kaplan, J., S. McCandlish, T. Henighan, T. Brown, B. Chess, R. Child, S. Gray, A. Radford, J. Wu, and D. Amodei. 2020. Scaling laws for neural language models. *arXiv preprint arXiv:2001.08361*. https://doi.org/10.48550/arXiv.2001.08361

[42] Kitchin, R. 2014. Big Data, new epistemologies and paradigm shifts. *Big Data & Society* 1 (1). DOI: 10.1177/2053951714528481;

[43] Kogan, V., S. Lavertu, and Z. Peskowitz. 2021. The democratic deficit in US education governance. *American Political Science Review* 115 (3): 1082–1089. DOI: 10.1017/S0003055421000162

[44] Kottonau, J. and C. Pahl-Wostl. 2004. Simulating political attitudes and voting behavior. *Journal of Artificial Societies and Social Simulation* 7 (4).

[45] Lewis, P., E. Perez, A. Piktus, F. Petroni, V. Karpukhin, N. Goyal, H. Küttler, M. Lewis, W. Yih, T. Rocktäschel, S. Riedel, and D. Kiela. 2020. Retrieval-augmented generation for knowledge-intensive NLP tasks. *Advances in Neural Information Processing Systems* 33: 9459–9474. DOI: 10.48550/arXiv.2005.11401

[46] Li, Hao, Ruoyuan Gong, and Hao Jiang. 2024. Political actor agent: Simulating legislative system for roll call votes prediction with large language models. *arXiv preprint arXiv:2412.07144*.

[47] Li, Lincan et al. 2024. Political-LLM: Large language models in political science. *arXiv preprint arXiv:2412.06864*.

[48] Liu, C. and S. Tang. 2023. The Limitations of Machine Learning in Conflict Prediction: A Discussion About the Predictive Validity of the ViEWS (机器学习在冲突预测方面的局限——基于对暴力预警系统的再检验与讨论). *World Economics and Politics* (12): 114-143+171-172.

[49] Liu, Xiao et al. 2023. AgentBench: Evaluating LLMs as agents. arXiv preprint arXiv:2308.03688.

[50] Lorè, N. and B. Heydari. 2023. Strategic behavior of large language models: game structure vs. contextual framing. *arXiv preprint arXiv. 2309.05898*. DOI: 10.48550/arXiv.2309.05898

[51] Marshall, J. 2019. The Anti-Democrat Diploma: How High School Education Decreases Support for the Democratic Party. *American Journal of Political Science* 63 (1): 67–83. DOI: 10.1111/ajps.12409;

[52] McDermott, R. 2002. Experimental methods in political science. *Annual Review of Political Science* 5 (1): 31-61. DOI: 10.1146/annurev.polisci.5.091001.170657

[53] Mi, J., C. Zhang, D. Li, and T. Lin. 2018. The Fourth Paradigm: The Transformation of Social Science Research Driven by Big Data (第四研究范式:大数据驱动的社会科学研究转型). *Academia Bimestris* (2):11-27. DOI:10.16091/j.cnki.cn32-1308/c.2018.02.003.

[54] Miao, Q. and F. Wang. 2024. *Artificial Intelligence for Science (AI4S): Frontiers and*

*Perspectives Based on Parallel Intelligence*. Springer.

[55] Mou, Xinyi et al. 2024. From individual to society: A survey on social simulation driven by large language model-based agents. arXiv preprint arXiv:2412.03563.

[56] Mutz, D. C. 2018. Status threat, not economic hardship, explains the 2016 presidential vote. *Proceedings of the National Academy of Sciences* 115(19): E4330-E4339. https://doi.org/10.1073/pnas.1718155115

[57] Ostrom, E. 1998. A behavioral approach to the rational choice theory of collective action: Presidential address, American Political Science Association 1997. *American Political Science Review* 92 (1): 1–22. DOI: 10.2307/2585925

[58] Paranjape, B., S. Lundberg, S. Singh, H. Hajishirzi, L. Zettlemoyer, and M. T. Ribeiro. 2023. ART: Automatic multi-step reasoning and tool-use for large language models. arXiv preprint arXiv:2303.09014. https://doi.org/10.48550/arXiv.2303.09014

[59] Park, J. S., S. O'Brien, J. Cai, C. J. Morris, M. R. Liang, and M. S. Bernstein. 2023. Generative agents: Interactive simulacra of human behavior. *Proceedings of the 36th Annual ACM Symposium on User Interface Software and Technology.* Article 2: 1-22. https://doi.org/10.1145/3586183.3606763

[60] Philpot, T. S. 2018. Race, gender, and the 2016 presidential election. *PS: Political Science & Politics* 51(4): 755-761. DOI:10.1017/S1049096518000896

[61] Popper, K. R. 1972. *Objective Knowledge: An Evolutionary Approach*. Oxford University Press.

[62] Qin, X. 2024. Intelligent Technologies and Methodological Transformations in the Social Sciences. *Chinese Political Science* 9 (1): 1-17. DOI: 10.1007/s41111-021-00197-y

[63] Sale, J., L. Lohfeld, and K. Brazil. 2002. Revisiting the Quantitative-Qualitative Debate: Implications for Mixed-Methods Research. *Quality and Quantity* 36: 43-53. DOI: 10.1023/A:1014301607592.

[64] Shapira, E., O. Madmon, R. Reichart, and M. Tennenholtz. 2024. Can Large Language Models Replace Economic Choice Prediction Labs? *arXiv preprint arXiv:2401.17435*. DOI: 10.48550/arXiv.2401.17435

[65] Shi, Freda et al. 2023. Large language models can be easily distracted by irrelevant context. In *International Conference on Machine Learning*, PMLR, p. 31210–31227.

[66] Simon, H. A. 1985. Human nature in politics: The dialogue of psychology with political science. *American Political Science Review* 79 (2): 293–304. DOI: 10.2307/1956650

[67] Squazzoni, F., W. Jager, and B. Edmonds. 2014. Social Simulation in the Social Sciences: A Brief Overview. *Social Science Computer Review* 32 (3): 279–294. DOI: 10.1177/0894439313512975;

[68] Sun, S., E. Lee, D. Nan, X. Zhao, W. Lee, and J. H. Kim. 2024. Random Silicon Sampling: Simulating Human Sub-Population Opinion Using a Large Language Model Based on Group-Level Demographic Information. *arXiv preprint arXiv:2402.18144*. https://doi.org/10.48550/arXiv.2402.18144

[69] Tang, S. 2012. Outline of a New Theory of Attribution in IR: Dimensions of Uncertainty and Their Cognitive Challenges. *The Chinese Journal of International Politics* 5 (3): 299–338. https://doi.org/10.1093/cjip/pos013

[70] Tang, S. 2018. Idea, action, and outcome: The objects and the tasks of social sciences (观念、行动和结果:社会科学的客体和任务). *World Economics and Politics* (5): 33-59+156.

[71] Tang, S., W. Held, O. Shaikh, J. Chen, Z. Zhang, and D. Yang. 2024. Idea, Action, and Outcome: The Objects and Tasks of Social Sciences. *Innovation in the Social Sciences* 2 (2): 123–170. DOI: 10.2139/ssrn.2790615.

[72] Wang, F. and Q. Miao. 2023. Novel Paradigm for AI-driven Scientific Research: From AI4S to Intelligent Science (人工智能驱动的科学研究新范式：从 AI4S 到智能科学). *Bulletin of Chinese Academy of Sciences* 38 (4): Article 2. https://doi.org/10.16418/j.issn.1000-3045.20230406002

[73] Wang, F. Y., X. Wang, L. Li, and L. Li. 2016. Steps toward parallel intelligence. *IEEE/CAA Journal of Automatica Sinica* 3 (4): 345–348. DOI: 10.1109/JAS.2016.7510067.

[74] Wang, H., T. Fu, Y. Du, W. Gao, K. Huang, Z. Liu, P. Chandak, S. Liu, P. Van Katwyk, A. Deac, A. Anandkumar, K. Bergen, C. P. Gomes, S. Ho, P. Kohli, J. Lasenby, J. Leskovec, T. Liu, A. Manrai, D. Marks, B. Ramsundar, L. Song, J. Sun, J. Tang, P. Veličković, M. Welling, L. Zhang, C. W. Coley, Y. Bengio and M. Zitnik. 2023. Scientific discovery in the age of artificial intelligence. *Nature* 620: 47–60. https://doi.org/10.1038/s41586-023-06221-2

[75] Wang, X., J. Wei, D. Schuurmans, Q. Le, E. Chi, S. Narang, A. Chowdhery, and D. Zhou. 2022. Self-Consistency Improves Chain of Thought Reasoning in Language Models. *arXiv preprint arXiv. 2203.11171*. DOI: 10.48550/arXiv.2203.11171

[76] Wang, Z. and S. Tang. 2020. Research on Forecasting Methods of Political Science—Taking Election Forecasting as an Example (政治科学预测方法研究——以选举预测为例). *CASS Journal of Political Science* (02): 52-64+126.

[77] Wei, J., X. Wang, D. Schuurmans, M. Bosma, B. Ichter, F. Xia, E. Chi, Q. Le, and D. Zhou. 2022a. Chain-of-thought prompting elicits reasoning in large language models. *Advances in neural information* 35: 24824-24837.

[78] Wei, J., Y. Tay, R. Bommasani, C. Raffel, B. Zoph, S. Borgeaud, D. Yogatama, M. Bosma, D. Zhou, D. Metzler, E. H. Chi, T. Hashimoto, O. Vinyals, P. Liang, J. Dean, and W. Fedus. 2022b. Emergent abilities of large language models. *arXiv preprint arXiv:2206.07682*. DOI: 10.48550/arXiv.2206.07682

[79] Wies, N., Y. Levine, and A. Shashua. 2024. The learnability of in-context learning. *Advances in Neural Information Processing Systems* 36.

[80] Xi, Z., W. Chen, X. Guo, W. He, Y. Ding, B. Hong, M. Zhang, J. Wang, S. Jin, E. Zhou, R. Zheng, X. Fan, X. Wang, L. Xiong, Y. Zhou, W. Wang, C. Jiang, Z. Zou, Y. Liu, Z. Yin, S. Dou, R. Weng, W. Cheng, Q. Zhang, W. Qin, Y. Zheng, X. Huang, and T. Gui. 2023. The Rise and Potential of Large Language Model Based Agents: A Survey. *arXiv preprint arXiv:2309.07864*. DOI: 10.48550/arXiv.2309.07864

[81] Xu, R., Y. Sun, M. Ren, S. Guo, R. Pan, H. Lin, L. Sun, and X. Han. 2024. AI for social

science and social science of AI: A survey. *Information Processing & Management* 61 (3): 103665. https://doi.org/10.1016/j.ipm.2024.103665

[82] Xu, Y., S. Wang, P. Li, F. Luo, X. Wang, W. Liu, and Y. Liu. 2023. Exploring Large Language Models for Communication Games: An Empirical Study on Werewolf. *arXiv preprint arXiv:2309.04658*. DOI: 10.48550/arXiv.2309.04658

[83] Yang, J., X. Wang, Y. Wang, Z. Liu, X. Li, and F. Wang. 2023. Parallel Intelligence and CPSS in 30 Years: An ACP Approach (平行智能与 CPSS:三十年发展的回顾与展望). *Acta Automatica Sinica* 49 (3): 614-634. DOI:10.16383/j.aas.c230015.

[84] Yang, Ziyi et al. 2024. OASIS: Open agent social interaction simulations with one million agents.*arXiv preprint arXiv:2411.11581*.

[85] Yu, Y., Y. Zhuang, J. Zhang, Y. Meng, A. Ratner, R. Krishna, J. Shen, and C. Zhang. 2024. Large language model as attributed training data generator: A tale of diversity and bias. *Advances in Neural Information Processing Systems* 36. https://doi.org/10.48550/arXiv.2306.15895

[86] Zhang, Y., Y. Li, L. Cui, D. Cai, L. Liu, T. Fu, X. Huang, E. Zhao, Y. Zhang, Y. Chen, L. Wang, A. T. Luu, W. Bi, F. Shi, and S. Shi. 2023. Siren's song in the AI ocean: a survey on hallucination in large language models. *arXiv preprint arXiv:2309.01219*. https://doi.org/10.48550/arXiv.2309.01219

[87] Zhao, Q., J. Wang, Y. Zhang, Y. Jin, K. Zhu, H. Chen, and X. Xie. 2023. CompeteAI: Understanding the Competition Dynamics in Large Language Model-based Agents. *arXiv preprint, arXiv:2310.17512*. DOI: 10.48550/arXiv.2310.17512

[88] Ziems, C., W. Held, O. Shaikh, J. Chen, Z. Zhang, and D. Yang. 2024. Can large language models transform computational social science? *Computational Linguistics* 50 (1): 237–291. https://doi.org/10.1162/coli_a_00502